%% file: simplefolds.tex
\documentclass[11pt]{article}
\usepackage{amsfonts,color,epsfig,latexsym,url}

\title{When Can You Fold a Map?}
\author{Esther M. Arkin%
          \thanks{Department of Applied Mathematics and Statistics,
            SUNY, Stony Brook, NY 11794-3600, USA,
            email: \{\texttt{estie}, \texttt{jsbm}\}%
            \texttt{@ams.sunysb.edu}.}
   \and Michael A. Bender%
          \thanks{Department of Computer Science, SUNY,
            Stony Brook, NY 11794-4400, USA, email:
            \{\texttt{bender}, \texttt{saurabh}, \texttt{skiena}\}%
            \texttt{@cs.sunysb.edu}.}
   \and Erik D. Demaine%
          \thanks{MIT Laboratory for Computer Science,
            200 Technology Square, Cambridge, MA 02139, USA,
            \{\texttt{edemaine}, \texttt{mdemaine}\}\texttt{@mit.edu}.}
   \and Martin L. Demaine\footnotemark[3]
   \and Joseph S. B. Mitchell\footnotemark[1]
   \and Saurabh Sethia\footnotemark[2]
   \and Steven S. Skiena
          \thanks{Computer Science, Oregon State University, 102 Dearborn Hall,
            Corvallis, OR 97331-3202, USA, \texttt{saurabh@cs.orst.edu}.}
}
\date{}

\advance \topmargin by -\headheight
\advance \topmargin by -\headsep
\evensidemargin \oddsidemargin
\let\latexcite=\cite
\def\cite{\nolinebreak\latexcite}
\let\latexref=\ref
\def\ref{\nolinebreak\latexref}

{\makeatletter \gdef\fps@figure{!htbp}}

\newtheorem{theorem}{Theorem}[section]
\newtheorem{lemma}[theorem]{Lemma}
\newtheorem{corollary}[theorem]{Corollary}

\def\problem#1:#2\par{\begin{description}\item[Problem:] \textbf{#1}\\#2%
                      \end{description}}

\makeatletter
\def\GrabProofArgument[#1]{ (#1): \egroup\ignorespaces}
\def\proof{\noindent\textbf\bgroup Proof%
           \@ifnextchar[{\GrabProofArgument}{: \egroup\ignorespaces}}

\makeatother

{\makeatletter
 \gdef\xxxmark{%
   \expandafter\ifx\csname @mpargs\endcsname\relax 
     \expandafter\ifx\csname @captype\endcsname\relax 
       \marginpar{xxx}
     \else
       xxx 
     \fi
   \else
     xxx 
   \fi}
 \gdef\xxx{\@ifnextchar[\xxx@lab\xxx@nolab}
 \long\gdef\xxx@lab[#1]#2{{\bf [\xxxmark #2 ---{\sc #1}]}}
 \long\gdef\xxx@nolab#1{{\bf [\xxxmark #1]}}
}

\let\epsilon=\varepsilon

\def\R{\mathbb{R}}
\def\complement#1{\mathrm{comp} (#1)}

\def\note#1:#2\par{\par\medskip\noindent\textbf{#1:} #2\par\medskip}

\newcommand{\estie}[1]{\textbf{Estie:} #1}
\newcommand{\michael}[1]{\textbf{Michael:} #1}
\newcommand{\erik}[1]{\textbf{Erik:} #1}
\newcommand{\marty}[1]{\textbf{Marty:} #1}
\newcommand{\joe}[1]{\textbf{Joe:} #1}
\newcommand{\saurabh}[1]{\textbf{Saurabh:} #1}
\newcommand{\steve}[1]{\textbf{Steve:} #1}

\renewcommand{\estie}[1]{}
\renewcommand{\michael}[1]{}
\renewcommand{\erik}[1]{}
\renewcommand{\marty}[1]{}
\renewcommand{\joe}[1]{}
\renewcommand{\saurabh}[1]{}
\renewcommand{\steve}[1]{}

\newcommand{\comment}[1]{{}}
\newcommand{\old}[1]{{}}



\def\isabstract{0}
\def\ispaper{1}
\let\mymode\ispaper     
\long\def\ab#1{\if\mymode\isabstract#1\fi}
\long\def\pa#1{\if\mymode\ispaper#1\fi}
\let\inappendix=\pa 

\begin{document}
\maketitle


\begin{abstract}
We explore the following problem: given a collection of creases on a
piece of paper, each assigned a folding direction of mountain or
valley, is there a flat folding by a sequence of simple folds?  There
are several models of simple folds; the simplest \emph{one-layer
simple fold} rotates a portion of paper about a crease in the paper by
$\pm 180^\circ$.  We first consider the analogous questions in one
dimension lower---bending a segment into a flat object---which lead to
interesting problems on strings.  We develop efficient algorithms for
the recognition of simply foldable 1D crease patterns, and
reconstruction of a sequence of simple folds.  Indeed, we prove that a
1D crease pattern is flat-foldable by any means precisely if it is by
a sequence of one-layer simple folds.

\old{
1. Why do mention "(weakly)" all the time? Usually, if it is not
mentioned as strongly np-complete problem, one presume that it is weak.
}

Next we explore simple foldability in two dimensions, and find a surprising
contrast: ``map'' folding and variants are polynomial, but slight
generalizations are NP-complete.  Specifically, we develop a
linear-time algorithm for deciding foldability of an orthogonal crease
pattern on a rectangular piece of paper, and prove that
it is (weakly) NP-complete to decide foldability of (1)
an orthogonal crease pattern on a orthogonal piece of paper, (2) a
crease pattern of axis-parallel and diagonal (45-degree) creases on a
square piece of paper, and (3) crease patterns without a
mountain/valley assignment.
\end{abstract}


\section{Introduction}
\label{Introduction}

\newenvironment{halfquote}
  {\list{}{\leftmargin=0.5\leftmargin \rightmargin=\leftmargin}\item[]}
  {\endlist}

\begin{quote}
\begin{quote}
\begin{quote}
The easiest way to refold a road map is differently.

\hfill --- Jones's Rule of the Road (M. Gardner \cite{Gardner-1983-mapfolding})
\end{quote}
\end{quote}
\end{quote}

\begin{figure}
\centerline{\psfig{file=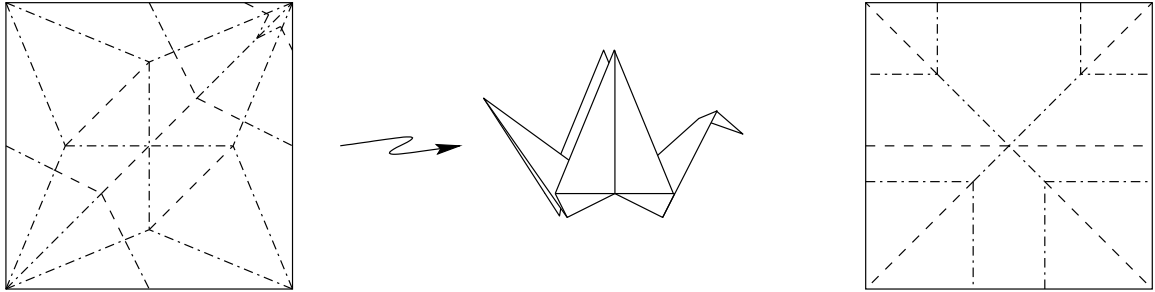}}
\caption{Sample crease patterns.  Left: the classic
  crane.   Right: pattern of Hull~\protect\latexcite{Hull-1994},
  which cannot be folded flat, for any mountain-valley assignment.}
%
\label{crease patterns}
\end{figure}

Perhaps the best-studied problem in origami 
%
%
mathematics is the characterization of flat-foldable crease patterns.
A crease pattern is a straight-edge embedding of a graph on a
polygonal piece of paper; a flat folding must fold along all of the
edges of the graph, but no more.  For example, two crease patterns are
shown in Figure~\ref{crease patterns}.  The first one folds flat into a
classic origami crane, whereas the second one cannot be folded flat
(unless the paper is allowed to pass through itself),
even though every vertex can be ``locally'' flat folded.

\old{The last bit was added by Erik in response to Saurabh's comment:
Hull's crease pattern is the simplest example of what?  There
surely are simpler examples of crease patterns that cannot
be folded flat even if we allow changing fold directions.
Is the speciality of Hull's example is that
it CAN be folded flat if the paper is allowed to pass through
itself?  In the first place, it's not clear to me what is meant
by allowing paper to pass through itself.}

The algorithmic version of this problem is to determine whether a
given crease pattern is flat-foldable.
\old{redundant:
Concisely, this is the problem of recognizing
flat-foldable crease patterns.}
The crease pattern may also have a direction of ``mountain'' or
``valley'' assigned to each crease, which restricts the way in which
the crease can be folded.  (Our figures adhere to the standard origami
convention that valleys are drawn as dashed lines and mountains are
drawn as dot-dashed lines.)

\begin{figure}
\centerline{\psfig{file=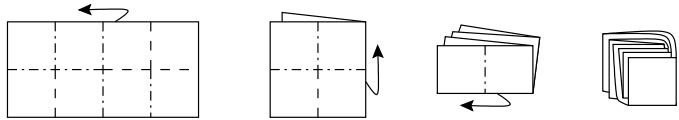}}
\caption{Folding a $2 \times 4$ map via a sequence of $3$ simple folds.}
\label{simple fold map}
\end{figure}

It is known that the general problem of deciding flat foldability of a
crease pattern is NP-hard~\cite{Bern-Hayes-1996}.
In this paper, we consider the important and very natural case of recognizing
crease patterns that arise as the result of flat foldings using
\emph{simple foldings}.  In this model, a flat folding is made by a
sequence of \emph{simple folds}, each of which folds one or more
layers of paper along a single line segment.
Figure \ref{simple fold map} shows an example of a simple folding.
As we define in
Section~\ref{Definitions}, there are different types of simple folds
(termed ``one-layer'', ``some-layers'', and ``all-layers''), depending
on how many layers of paper are required or allowed to be folded along
a crease.

Unsurprisingly, not every flat folding can be achieved by a simple folding.
For example, the crane in Figure~\ref{crease patterns} (top) cannot
be made by a simple folding.  In particular, there is no uniformly
mountain or valley segment that could serve as the first simple fold.
Also, the hardness gadgets of \cite{Bern-Hayes-1996} require nonsimple folds
which allow the paper to curve during folding \cite{Demaine-Mitchell-2001}.
Thus, the complexity of general flat foldability has no direct connection to
simple foldability.

The problem we study in this paper is that of determining whether a
given crease pattern (usually with specified mountain and valley
assignments) can be folded flat by a sequence of simple folds, and if
so, to construct such a sequence of folds.

Several of our results are based on the special case in which the
creases in the piece of paper are all parallel to one another.  This
case is equivalent to a \emph{one-dimensional} folding
problem of folding a line segment (``paper'') according to a set of
prescribed crease points (possibly labeled ``mountain'' or
``valley'').  We will therefore refer to this special case, which has
a rich structure of its own, as the ``1D'' case to distinguish it
from the general 2D problem.  In contrast to the 2D problem, we show
that 1D flat foldability is equivalent to 1D simple foldability.

\paragraph{Motivation.}
In addition to its inherent interest in the mathematics of origami,
our study is motivated by applications in sheet metal and paper
product manufacturing, where one is interested in determining whether a
given structure can be manufactured using a given machine. (See references
cited below.) 
%
%
While origamists can develop particular skill in
performing nonsimple folds to make beautiful artwork, practical
problems of manufacturing with sheet goods require simple and
constrained folding operations.  Our goal is to develop a first suite
of results that may be helpful towards a fuller algorithmic 
understanding of the
several manufacturing problems that arise, e.g., in making
three-dimensional cardboard and sheet-metal structures.

\paragraph{Related Work.}
Our problems are related to the classic combinatorics questions of
\emph{map folding}~\cite{Gardner-1983-mapfolding, Lunnon-1971}.
These questions ask for the \emph{number} of different flat foldings of a
particular crease pattern, namely an $m \times n$ rectangular grid, either by a
sequence of simple folds or by a general flat folding.  Two foldings are
usually considered ``different'' in this context if they differ in
the total order of the faces in the folding.
These questions have been studied extensively
\cite{Gardner-1983-mapfolding, Lunnon-1971}, particularly in the
one-dimensional ($1 \times n$) case
\cite{DiFrancesco-2000, Koehler-1968, Lunnon-1968, Touchard-1950},
but remain largely unsolved.
In contrast with these combinatorial questions, we study the algorithmic
complexity of the decision problems, and for more general crease patterns.

The mathematical and algorithmic problems arising in the study of flat
origami have been examined by several researchers, e.g., 
Hull~\cite{Hull-1994}, Justin~\cite{Justin-1994},
Kawasaki~\cite{Kawasaki-1989a}, and Lang~\cite{Lang-1996}.  Of particular
relevance to our work is the paper by
Bern and Hayes~\cite{Bern-Hayes-1996}, which shows that
the general problem of deciding flat foldability of a crease
pattern is strongly NP-hard.
Demaine~et~al.~\cite{Demaine-Demaine-Mitchell-2000} used
computational geometry techniques to show that any polygonal
(connected) silhouette can be obtained
by simple folds from a rectangular piece of paper.

\pa{
Our model of simple folding is also closely related to ``pureland
origami'', a restriction introduced by Smith~\cite{Smith-1976,
Smith-1980-1988-1993}.  Pureland folds include simple folds, but they
also allow paper to be ``tucked'' into pockets, as well as ``opened
up'' into three dimensions provided that no creases are made during
the process.
}

\old{comment from Bob Connelly from email 5/24/99:

I am not an expert on this sort of problem, but I remember as a kid that
there was a Martin Gardner column in the Scientific American about this sort
of problem.  A good place to start might be Chapter 7 in Martin's book
"Wheels, Life and Other Mathematical Amusements" (1983) Freeman.  He has an
addendum to the main article and references to other papers.  

His quotation at the start of the chapter "The easiest way to refold a
roadmap is differently".

Erik added:

Here is a more complete reference:

    Martin Gardner.  Combinatorics of Paper-Folding.
    Wheels, Life, and other Mathematical Amusements.
    pp. 60-73. 
}

There has been quite a bit of work on the related problems of
manufacturability of sheet metal parts (see e.g.\ \cite{Wang-1997})
and folding cartons (see e.g.\ \cite{Lu-Akella-2000}).  Existing CAD/CAM
techniques (including BendCad and PART-S) rely on worst-case
exponential-time state space searches (using the A$^*$ algorithm). In
general, the problem of bend sequence generation is a challenging (and
provably hard~\cite{Arkin-Fekete-Mitchell-2003})
coordinated motion planning problem.  For example, Lu and
Akella~\cite{Lu-Akella-2000} utilize a novel configuration-space
formulation of the folding sequence problem for folding cartons using
fixtures; their search, however, is still worst-case exponential time.
Our work differs from the prior work on sheet metal and cardboard
bending in that the structures we are folding are ultimately ``flat''
in their folded states (all bend angles in the input crease pattern
are $\pm 180^\circ$, according to a mountain-valley assignment that is part
of the input crease pattern).  Also, we are concerned only with the
feasibility of the motion of the (stiff) material that is being
folded---does it collide with itself during the folding motion?  We are not
addressing here the issues of reachability by the tools that perform
the folding.  As we show, even with the restrictions that come with 
the problems we study, there is a rich mathematical and algorithmic
theory of foldability.

\old{
(3) I'd like to see more motivation. Srinivas Akella at RPI
has done something with box folding-- how close is this problem
to your results?
}
%

\paragraph{Summary of Our Results.}
We develop a variety of new algorithmic results
(see Table \ref{results}):

\begin{enumerate}
\item[(1)]
We analyze the 1D one-layer and some-layers cases, giving
a full characterization of flat-foldability and an $O(n)$ algorithm
for deciding foldability and producing a folding sequence, if one
exists.

\item[(2)]
We analyze the 1D all-layers case as a ``string folding''
problem.  In addition to a simple $O(n^2)$ algorithm, we give an
algorithm utilizing suffix trees that requires time linear in the bit
complexity of the input, and a randomized algorithm with expected
$O(n)$ running time.

\item[(3)] 
We give an algorithm for deciding simple foldability of orthogonal crease
patterns on a rectangular piece of paper%
\footnote{Throughout this paper, the notions of ``orthogonal'' and
  ``rectangular'' implicitly require axis-parallelism
  with a common set of axes.}
(the ``map folding problem''),
in the one-, some-, and all-layers cases, based on and with the same running
times as our 1D results.

\item[(4)]
We prove that it is (weakly) NP-complete to decide simple foldability of
an orthogonal crease pattern on a piece of paper that is more general
than a rectangle: a simple orthogonal polygon.

\item[(5)] 
We also prove that it is (weakly) NP-complete to
decide simple foldability of a square piece of paper with a crease pattern
that includes {\em diagonal} creases (angled at $45^\circ$), in addition to
axis-parallel creases.

\item[(6)] 
We show that it is (weakly) NP-complete to decide
simple foldability of an orthogonal piece of paper having a crease pattern
for which no mountain-valley assignment is given.

\end{enumerate}

\begin{table}[t]
  \centering
  \small
  \advance \tabcolsep -1pt
  \def\notequiv{\not\equiv}
  \begin{tabular}{|lll|lllllll|}
  \multicolumn{1}{l}{}&&\multicolumn{1}{l}{}
     & \multicolumn{7}{c}{\bf Model of folding} \\ \cline{4-10}
  \multicolumn{1}{l}{}
     &&& All-layers   && Some-layers  && One-layer    && General      \\
  \multicolumn{1}{l}{\bf Dim.} & \bf Paper & \bf Creases
     & simple folds && simple folds && simple folds && flat folding \\
  \cline{1-3}
       \cline{4-4}     \cline{6-6}     \cline{8-8}     \cline{10-10}
  1D & &
     & $O(n)$ rand.    & $\notequiv$
                       & $O(n)$ & $\equiv$ & $O(n)$ & $\equiv$ & $O(n)$ \\
     &&& $O(n \lg n')$ det.&&&&&&\\
  \hline
  2D & Rect & Ortho
     & $O(n)$ rand.    & $\notequiv$ & $O(n)$ & $\notequiv$ & $O(n)$
                           & $\notequiv$ & Open \cite{Edmonds-1997-personal} \\
     &&& $O(n \lg n')$ det.&&&&&&\\
  \hline
  2D & Ortho & Ortho
     & Weakly      && Weakly      && Weakly      && Open \\
  or & Square & Ortho${} + 45^\circ$
     & NP-complete && NP-complete && NP-complete && \\
  \hline
  2D & Square & General &&&&&&& Strongly \\
     &        &         &&&&&&& NP-hard \cite{Bern-Hayes-1996} \\
  \hline
  \end{tabular}
  \caption{Summary of the complexities of deciding flat foldability by various
    models of simple folds, and by general flat foldings.
    The symbols $\equiv$ and $\notequiv$ denote equivalences and
    nonequivalences between certain models.
    The abbreviations ``rand.''\ and ``det.''\ denote randomized and
    deterministic algorithms, respectively.}
  \label{results}
\end{table}

Note that our hardness results do not strengthen those of
\cite{Bern-Hayes-1996}, because deciding simple foldability is different
from deciding flat foldability.

\old{
The rest of this paper is organized as follows.  We begin in Section
\ref{Definitions} with more formal definitions of the problem.
Then in Section~\ref{1D Some Layers} we study the one-layer
and some-layers folding problems in one dimension, where we prove the
equivalence of flat foldability, some-layer simple foldability, and
one-layer simple foldability in 1D, and give a linear-time algorithm
for detection.
In Section~\ref{1D All Layers} we examine the all-layers problem in
one dimension, giving two different efficient algorithms (one
deterministic, one randomized) for determining foldability.
We generalize these results to two dimensions for the special case of
orthogonal folds on a rectangle in Section~\ref{ortho-2D}.
\comment{We generalize further to two dimensions and unrestricted folding
directions in Section~\ref{unrestricted-2D}.}
Finally, in Section~\ref{Unrestricted 2D}
we show that slight generalizations of this problem are NP-hard.
}

\section{Definitions}
\label{Definitions}

We are concerned with foldings in one and two dimensions, although several of
our definitions and results extend to higher dimensions.  A one-dimensional
piece of paper is a (line) \emph{segment} in $\R^1$.  A two-dimensional piece
of paper is a (connected) \emph{polygon} in $\R^2$, possibly with holes.  In
both cases, the paper is folded through one dimension higher than the object;
thus, segments are folded through $\R^2$ and polygons are folded through
$\R^3$.  \emph{Creases} have one less dimension; thus, a crease is a point on a
segment and a line segment on a polygon.

\begin{figure}
\centerline{\psfig{file=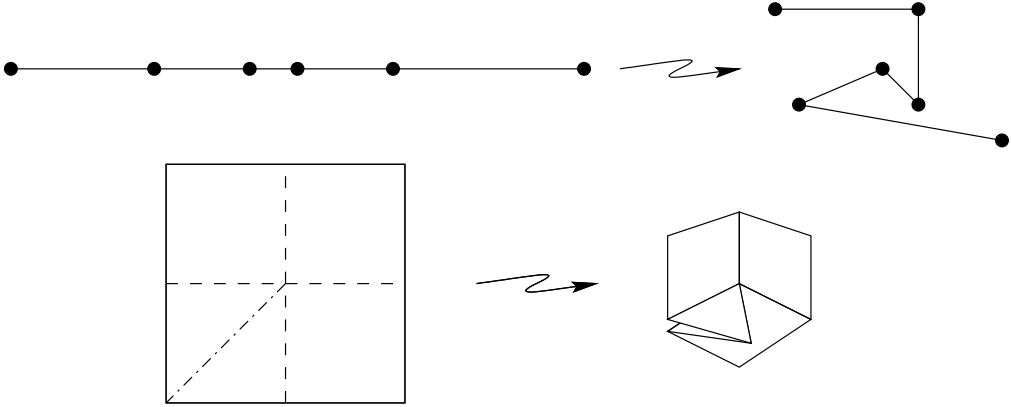}}
\caption{\label{nonflat folding}
  Sample nonflat foldings in one and two dimensions.}
\end{figure}

A \emph{crease pattern} is a collection of creases on the piece of paper, no
two of which intersect except at a common endpoint.  A \emph{folding} of a
crease pattern is an isometric embedding of the piece of paper, bent along
every crease in the crease pattern (and not bent along any segment that is not
a crease).  In particular, each facet of
paper must be mapped to a congruent copy, the connectivity between facets must
be preserved, and the paper cannot cross itself, although multiple layers of
paper may touch.
See Figure~\ref{nonflat folding}.

\begin{figure}
\centerline{\psfig{file=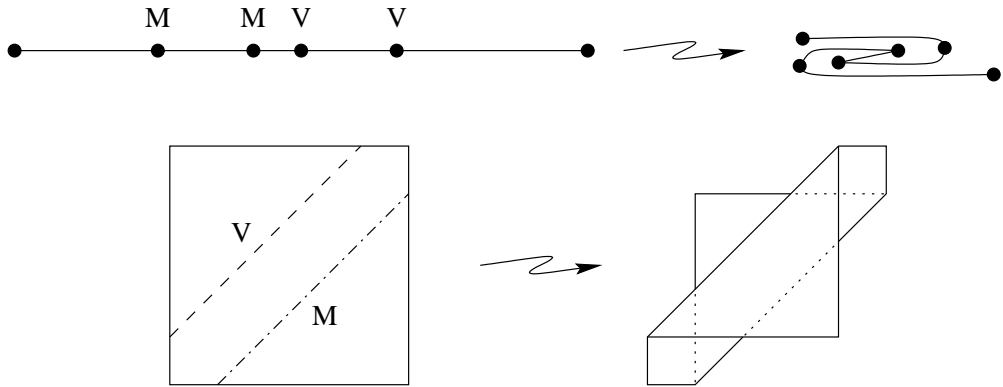}}
\caption{\label{flat folding}
  Sample flat foldings in one and two dimensions.
  Mountains and valleys are denoted by M's and V's, respectively.}
\end{figure}

A \emph{flat folding} has the additional property that it lies in the same
space as the unfolded piece of paper.  That is, a flat folding of a segment
lies in $\R^1$, and a flat folding of a polygon lies in $\R^2$.  In reality,
there can be multiple layers of paper at a point, so the folding really
occupies a finite number of infinitesimally close copies of $\R^1$ or $\R^2$.
See Figure~\ref{flat folding}.  More formally, a flat folding can be specified
by a function mapping the vertices to their folded positions, together with a
partial order of the facets of paper that specifies their overlap order
\cite{Bern-Hayes-1996,Hull-1994,Lang-1996}.
For each pair of facets of the crease pattern that fold to overlapping
polygons, this partial order must specify which facet is layered above the
other.

If we orient the piece of paper to have a top and bottom side, we can talk
about the \emph{direction} of a crease in a flat folding.  A \emph{mountain}
brings together the bottom sides of the two adjacent facets of paper, and a
\emph{valley} brings together the top sides.  A \emph{mountain-valley
assignment} is a function from the creases in a crease pattern to $\{M, V\}$.
This is the labeling shown in Figure~\ref{flat folding}.  Together, a crease
pattern and a mountain-valley assignment form a \emph{mountain-valley pattern}.

This paper is concerned with the following general question:

\problem Simple Folding:
Given a mountain-valley pattern, is there a simple folding satisfying the
specified mountains and valleys?  If so, construct such a simple folding.

\begin{figure}
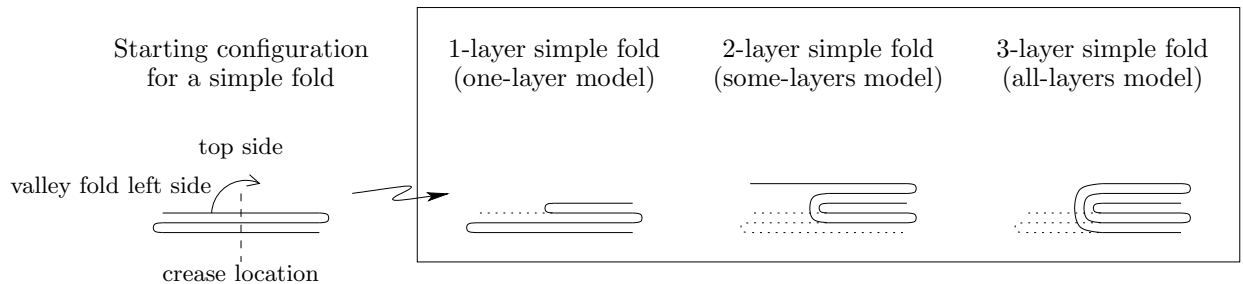

  \centering
  \hspace{10pt}
  \input various_layers_1d.pstex_t
  \caption{Illustration of a simple fold in 1D,
    which is specified by a crease location, a notion of the ``top side'',
    how many of the top layers are folded (here ranging from $1$ to $3$),
    whether the fold is mountain or valley, and
    whether the left or right side is folded.}
  \label{various layers 1D}
\end{figure}

There are three natural versions of this problem, depending on the type of
``simple folds'' allowed.
In general, a \emph{simple folding} is a sequence of
simple folds.  Each simple fold takes a flat-folded piece of paper, and folds
it into another flat folding using additional creases.  We distinguish
three types of simple folds: one-layer, all-layers, and some-layers.
Refer to Figure \ref{various layers 1D}.

We begin by defining in 1D the most general type of simple fold,
the some-layers simple fold.
A \emph{some-layers simple fold} is specified by
(1)~an orientation of the folded piece of paper to specify a \emph{top side},
(2)~an externally visible crease (point) on the top side of
    the folded piece of paper,
(3)~the number $\ell$ of layers to be folded, and
(4)~the orientation of the fold, mountain or valley,
    relative to which side is the top.
Such a fold newly creases the piece of paper at $\ell$ points: at the specified
crease on the topmost layer and at the $\ell-1$ creases (points) immediately
below.
If we locally color the piece of paper near the new creases, blue to the left
of the creases and red to the right of the creases, and propagate this
coloring, we should obtain a partition of the piece of paper into two (not
necessarily connected) components.  If we find a conflict that some paper
should be colored simultaneously red and blue, the simple fold is not valid.
Otherwise, the execution of the simple fold corresponds to continuously
rotating the blue portion of paper by $180^\circ$ around the crease point,
either clockwise or counterclockwise according to whether the fold is
valley or mountain.
During this rotation, both the red and blue portions of the paper remain rigid.
If the paper self-intersects during this rotation, the simple fold is invalid.

Some-layers simple folds are most general in the sense that any number $\ell$
of layers can be folded at once.  A \emph{one-layer simple fold} is the special
case in which $\ell = 1$.  An \emph{all-layers simple fold} is the special case
in which $\ell$ is the entire number of layers coinciding at the specified
crease point.

\begin{figure}
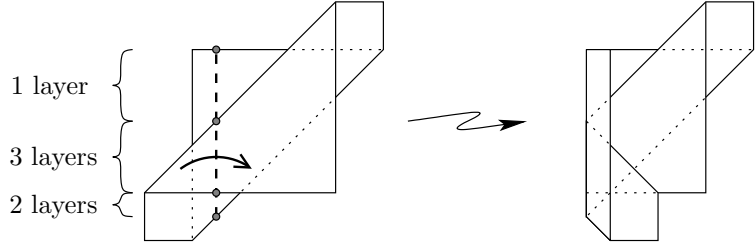

  \centering
  \input various_layers_2d.pstex_t
  \caption{Illustration of a simple fold in 2D
    (starting from Figure \protect\ref{flat folding}, right),
    where we must specify the number of folded layers
    for each portion of the crease,
    because this number can vary along a single fold.}
  \label{various layers 2D}
\end{figure}

In 2D, the situation is more complicated because the number of layers folded
can vary along the crease segment.  See Figure \ref{various layers 2D} for an
example.  Thus, a some-layers simple fold must specify the desired number of
layers for each portion of the crease, for a specified subdivision of the
crease segment into portions.  We construct the new creases that result from
this fold by copying each portion of the crease to the specified number of
layers below the topmost layer.  Then, as before, we color the piece of paper,
verify that this coloring is consistent (in particular, verifying
that the assignment of layers was valid), and rotate the blue portion of paper,
barring self-intersection.

In this more general setting, a one-layer simple fold is the special case of
folding only one layer along the entire crease.  An all-layers simple fold is
the special case of folding all layers for each portion of the crease.
The simple fold in Figure \ref{various layers 2D} is an example that cannot be
made a one-layer simple fold, and indeed, cannot be modified to use any smaller
number of layers at any point.  This fact can be verified by attempting to fold
such a piece of paper in practice, or by checking that the resulting red-blue
coloring is invalid.

\begin{figure}
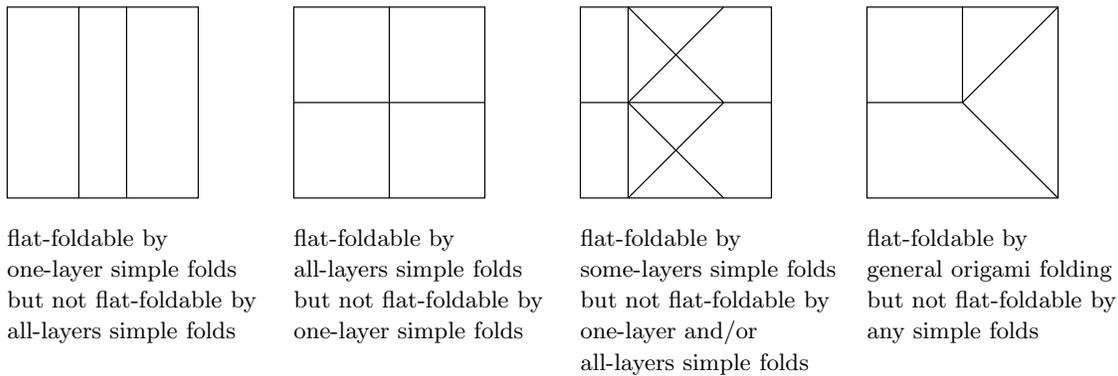

  \centering
  \input model_power.pstex_t
  \caption{Crease patterns illustrating the unique power of each model of
    simple folding, and the limitations compared to general flat folding.}
  \label{model power}
\end{figure}

Figure \ref{model power} further illustrates the differences among the three
models of simple folding, and their limitations with respect to general flat
folding, by giving examples of crease patterns that can be folded with one
model but not the others.  Of course, flat foldability by one-layer simple
folds or by all-layers simple folds implies flat foldability by some-layers
simple folds, which in turn implies flat foldability in general.
The examples in the figure prove that no other general implications hold.

\erik{I'm currently omitting the models that allow unfolding, in which the goal
is simply to make all the creases and not to actually fold along all the
creases at once.  This keeps the paper more focused.  But I'm certainly willing
to put the models in if we come up with algorithms or complexity results for
recognizing these classes.
}

\section{1D: One-Layer and Some-Layers}
\label{1D Some Layers}

\begin{figure}
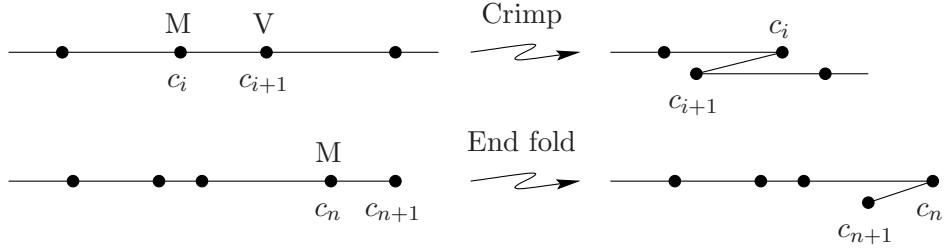

\centerline{\input local_ops.pstex_t}
\caption{\label{local ops}
  The two local operations for one-dimensional one-layer folds.}
\end{figure}

This section is concerned with the 1D one-layer simple-fold problem.
We will prove the surprising result that we only need to search for one of two
local operations to perform.  The two operations are called \emph{crimps} and
\emph{end folds}, and are shown in Figure~\ref{local ops}.

More formally, let $c_1, \dots, c_n$ denote the creases on the segment,
oriented so that $c_i$ is left of $c_j$ for $i < j$.
Let $c_0$ [$c_{n+1}$] denote the left [right] end of the segment.
Despite the ``$c$'' notation (which is used for convenience), $c_0$ and
$c_{n+1}$ are not considered \emph{creases}; instead they are called the
\emph{ends}.

First, a pair $(c_i, c_{i+1})$ of consecutive creases is \emph{crimpable} if
$c_i$ and $c_{i+1}$ have opposite directions and
$$|c_{i-1} - c_i| \geq |c_i - c_{i+1}| \leq |c_{i+1} - c_{i+2}|.$$
\emph{Crimping} such a pair corresponds to folding $c_i$ and then folding
$c_{i+1}$, using one-layer simple folds.

Second, $c_0$ is a \emph{foldable end} if $|c_0 - c_1| \leq |c_1 - c_2|$, and
$c_{n+1}$ is a \emph{foldable end} if $|c_{n-1} - c_n| \geq |c_n - c_{n+1}|$.
\emph{Folding} such an end corresponds to performing a one-layer simple fold at
the nearest crease (crease $c_1$ for end $c_0$, and crease $c_n$ for end
$c_{n+1}$).

We claim that one of the two local operations exists in any flat-foldable 1D
mountain-valley pattern.  We claim further that an operation exists for any
pattern satisfying a certain ``mingling property''.  Specifically, a 1D
mountain-valley pattern is called \emph{mingling} if for every sequence $c_i,
c_{i+1}, \dots, c_j$ of consecutive creases with the same direction, either
  \begin{enumerate}
  \item $|c_{i-1} - c_i| \leq |c_i - c_{i+1}|$; or
  \item $|c_{j-1} - c_j| \geq |c_j - c_{j+1}|$.
  \end{enumerate}
We call this the mingling property because, for maximal sequences of
consecutive creases with the same direction, it says that there are folds of
the opposite direction nearby.  In this sense, the mountain-valley pattern is
``crowded'' and the mountains and valleys must ``mingle'' together.

First we show that mingling mountain-valley patterns include flat-foldable
patterns:

\begin{lemma} \label{flat-foldable implies mingling}
Every flat-foldable 1D mountain-valley pattern is mingling.
\end{lemma}

\begin{figure}
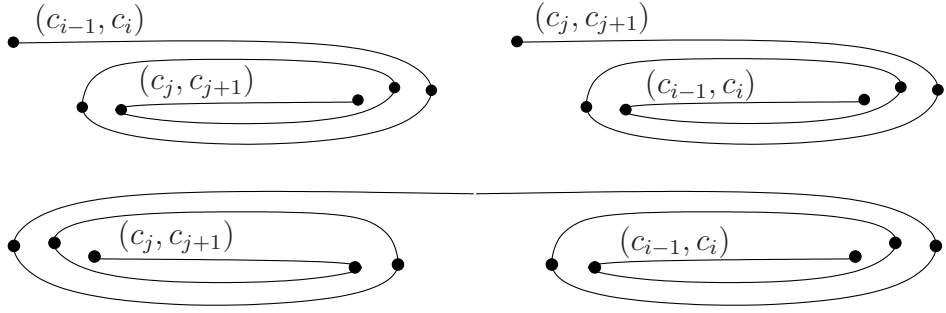

\centerline{\input spirals.pstex_t}
\caption{\label{spirals} The innermost edge of a spiral cannot be longer than
  the adjacent edge, in contrast to the outermost edge which can be arbitrarily
  long.}
\end{figure}

\begin{proof}
Consider a flat folding of a mountain-valley pattern, and let $c_i, \dots, c_j$
be consecutive creases with the same direction.
The portion $c_{i-1}, \dots, c_{j+1}$ of the segment can be in one of three
configurations (see Figure~\ref{spirals}):
  \begin{enumerate}
  \item The portion forms a ``spiral'' with $(c_{i-1}, c_i)$ being the
        outermost edge of the spiral, and $(c_j, c_{j+1})$ being the
        innermost; or
  \item The portion forms a ``spiral'' with $(c_j, c_{j+1})$ being the
        outermost edge of the spiral, and $(c_{i-1}, c_j)$ being the
        innermost; or
  \item The portion forms two ``spirals'' connected by a common outermost
        edge and with $(c_{i-1}, c_i)$ and $(c_j, c_{j+1})$ being the
        two innermost edge.
  \end{enumerate}

Now if $|c_{i-1} - c_i| > |c_i - c_{i+1}|$, then $(c_{i-1}, c_i)$ cannot be the
innermost edge of a spiral, or else $(c_{i-1}, c_i)$ would penetrate through
$c_{i+1}$.  Similarly, if $|c_{j-1} - c_j| > |c_j - c_{j+1}|$, then $(c_{j-1},
c_j)$ cannot be the innermost edge of the spiral.  Because in all three
configurations above we must have at least one of $(c_{i-1}, c_i)$ and
$(c_j, c_{j+1})$ as innermost, we cannot have both inequalities true.
\end{proof}

Next we show that having the mingling property suffices to imply the existence
of a single crimpable pair or foldable end.

\begin{lemma} \label{mingling implies fold exists}
Any mingling 1D mountain-valley pattern has either a crimpable pair or a
foldable end.
\end{lemma}

\begin{proof}
Let $i$ be maximum such that $c_1, \dots, c_i$ all have the same direction.
By the mingling property, either $|c_0 - c_1| \leq |c_1 - c_2|$
or $|c_{i-1} - c_i| \geq |c_i - c_{i+1}|$.
In the former case, $c_0$ is a foldable end, so we have the desired result.
A generalization of the latter case is that we have $c_i, \dots, c_j$
all with the same orientation, and $|c_{j-1} - c_j| \geq |c_j - c_{j+1}|$.
If $j = n$, then $c_{n+1}$ is a foldable end, so we have the desired result.
Otherwise, let $k$ be maximum such that $c_{j+1}, \dots, c_k$ all have the same
direction.  By the mingling property, either $|c_j - c_{j+1}| \leq |c_{j+1} -
c_{j+2}|$ or $|c_{k-1} - c_k| \geq |c_k - c_{k+1}|$.  In the former case,
$(c_j, c_{j+1})$ is a crimpable pair, so we have the desired result.
In the latter case, induction applies.
\end{proof}

\begin{figure}
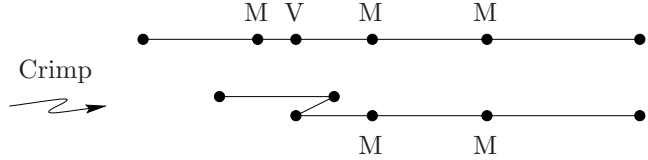

\centerline{\input mingling_unfoldable.pstex_t}
\caption{\label{mingling unfoldable}
  A mingling mountain-valley pattern that when crimped is no longer mingling
  and hence not flat-foldable.
  Indeed, the original mountain-valley pattern is not flat-foldable.}
\end{figure}

Ideally, we could show at this point that performing either of the two local
operations preserves the mingling property, and hence a mountain-valley pattern
is mingling precisely if it is flat-foldable.  Unfortunately this is false, as
illustrated in Figure~\ref{mingling unfoldable}.  Instead, we must prove that
flat foldability is preserved by each of the two local operations; in other
words, if we treat the folded object from a single crimp as a new segment, it
is flat-foldable.

\begin{lemma} \label{folding end preserves foldability}
Folding a foldable end preserves flat foldability.
\end{lemma}

\begin{proof}
This is obvious because folding a foldable end is equivalent to chopping off a
portion of the segment.  Thus, if we take a flat folding of the original
pattern, remove that portion of the segment, and double up the number of layers
for the adjacent portion of the segment, we have a flat folding of the new
object.
\end{proof}

\begin{lemma} \label{crimping preserves foldability}
Crimping a crimpable pair preserves flat foldability.
\end{lemma}

\begin{figure}
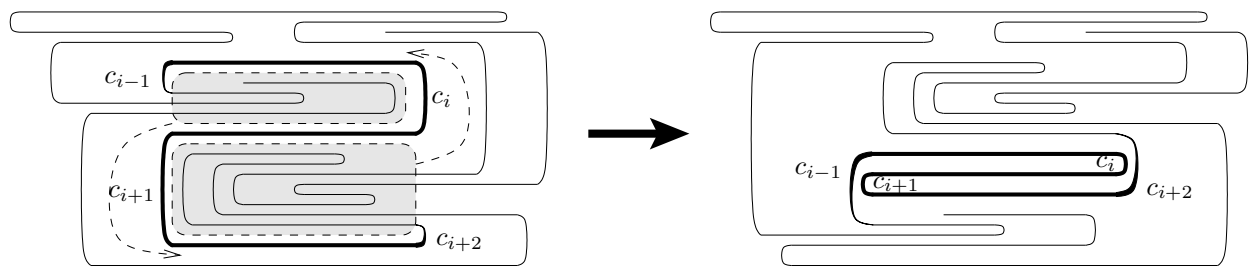

\centerline{\hspace{10pt}\input moving_stuff.pstex_t}
\caption{\label{moving stuff}
  Moving layers of paper out of the zig-zag formed by a crimp
  $(c_i, c_{i+1})$, highlighted in bold.}
\end{figure}

\begin{proof}
Let $(c_i, c_{i+1})$ be a crimpable pair, and assume by symmetry that $c_i$ is
a mountain and $c_{i+1}$ is a valley.  Consider a flat folding $F$ of the
original segment, such as the one in Figure~\ref {moving stuff} (left).
We orient our view to regard the segment $(c_i, c_{i+1})$
as flipping over during the folding, so that the remainder of the (unfolded)
segment keeps the same orientation.  Thus, $(c_{i-1}, c_i)$ is above
$(c_i, c_{i+1})$ which is above $(c_{i+1}, c_{i+2})$.
Now suppose that $F$ places some layers of paper
in between $(c_i, c_{i+1})$ and $(c_{i+1}, c_{i+2})$.  Then these layers of
paper can be moved to immediately above $(c_{i-1}, c_i)$, because $(c_{i-1},
c_i)$ is at least as long as $(c_i, c_{i+1})$, and hence there are no barriers
closer than $c_i$.  See Figure~\ref{moving stuff}.  Similarly, we move material
between $(c_i, c_{i+1})$ and $(c_{i-1}, c_i)$ to immediately below $(c_{i+1},
c_{i+2})$.  In the end, we have a flat folding of the object obtained from
making the crimp $(c_i, c_{i+1})$.
\end{proof}

Combining all of the previous results, we have the following:

\begin{theorem} \label{big some-layers result}
Any flat-foldable 1D mountain-valley pattern can be folded by a sequence of
crimps and end folds.
\end{theorem}

\begin{proof}
By Lemma \ref{flat-foldable implies mingling}, the pattern is mingling, and
hence by Lemma \ref{mingling implies fold exists} we can find a crimpable pair
or a foldable end.  Making this fold preserves flat foldability by Lemmas
\ref{folding end preserves foldability} and \ref{crimping preserves
foldability}, so by induction the result holds.
\end{proof}

A particularly interesting consequence of this theorem is the following
connection to general flat foldability:

\begin{corollary}
The following are equivalent for a 1D mountain-valley pattern $P$:
  \begin{enumerate}
  \item $P$ has a flat folding.
  \item $P$ has a some-layers simple folding.
  \item $P$ has a one-layer simple folding.
  \end{enumerate}
\end{corollary}

One-dimensional flat foldability has been studied extensively
in the combinatorial context
\cite{DiFrancesco-2000, Koehler-1968, Lunnon-1968, Touchard-1950},
but primarily for the simple crease pattern in which the distances
between consecutive creases are identical.
A structure similar to ours, in particular highlighting the importance of
crimps, is hinted at by Justin \cite[Section~6.1]{Justin-1994},
though it is not followed through algorithmically.

Finally, we show that Theorem \ref{big some-layers result} leads to a
simple linear-time algorithm:

\begin{theorem}
The 1D one-layer and some-layers simple-fold problems can be solved in $O(n)$
worst-case time on a machine supporting arithmetic on the input lengths.
\end{theorem}

\begin{proof}
First note that it is trivial to check in constant time whether a pair of
consecutive folds form a crimp or whether an end is foldable.  We begin by
testing all such folds, and hence in linear time have a linked list of all
possible folds at this time.  We also maintain reverse pointers from each
symbol in the string to the closest relevant possible fold.  Now when we make a
crimp or an end fold, only a constant number of previously possible folds can
no longer be possible, and a constant number of previously impossible folds can
be newly possible.  These folds can be discovered by examining a constant-size
neighborhood of the performed fold.  We remove the old folds from the list of
possible folds, and add the new folds to the list.  Then we perform the first
fold on the list, and repeat the process.  By Theorem \ref{big some-layers
result}, if the list ever becomes empty, it is because the mountain-valley
pattern is not flat-foldable.
\end{proof}

\section{1D: All-Layers Simple Folds}
\label{1D All Layers}

The 1D all-layers simple-fold problem can be cast as an interesting
``string folding'' problem.  (This folding problem is not to be confused with
the well-known protein/string folding problem in biology
\cite{Crescenzi-Goldman-Papadimitriou-Piccolboni-Yannakakis-1998}.)  The input
mountain-valley pattern can be thought of as a string of lengths
interspersed with mountain and valley creases.  Specifically, we will assume
that the input lengths are specified as integers or equivalently rational
numbers.  (Irrational numbers can be replaced by close rational
approximations, provided the sorted order of the lengths is preserved.)

Thus, an input string is of the form $\ell_0$ $d_1$ $\ell_1$ $d_2$ $\cdots$
$d_{n-1}$ $\ell_{n-1}$ $d_n$ $\ell_n$, where each $d_i \in \{M, V\}$
specifies the direction of the $i$th crease $c_i$, and each $\ell_i$ is a
positive rational number specifying the distance between adjacent creases $c_i$
and $c_{i+1}$.
We call each $d_i$ and $\ell_i$ a \emph{symbol} of the string.
It will be helpful to introduce some more uniform notation for symbols.
For a string $S$ of length $N = 2 n + 1$,
we denote the $i$th symbol by $S[i]$, where $1 \leq i \leq N$.

When we make an all-layers simple fold, we cannot ``cover up'' a crease
except with a matching crease (which when unfolded is in fact the other
direction), because otherwise this crease will be impossible to fold later.  To
formalize this condition, we define the \emph{complement} of symbols in the
string: $\complement{\ell_i} = \ell_i$, $\complement{M} = V$, and
$\complement{V} = M$.
For each even index $i$, at which $S[i] = d_{i/2} \in \{M,V\}$,
we define the \emph{fold at position $i$} to be the all-layers simple fold
of the corresponding crease $c_{i/2}$.
We call this fold \emph{allowable} if
$S[i-x] = \complement{S[i+x]}$ for all $1 \leq x \leq \min(i-1, N-i)$, except
that $S[1]$ and $S[N]$ (the end lengths) are allowed to be shorter than their
complements.

\begin{lemma} \label{allowable fold preserves foldability}
A mountain-valley pattern can be folded by a sequence of all-layers simple
folds precisely if there is an allowable fold, and the result after performing
that fold has an allowable fold, and so on, until all creases of the segment
have been folded.
\end{lemma}

\begin{proof}
Performing an all-layers simple fold that is not allowable forbids us from
all-layers simple folding certain creases, and hence the resulting segment
cannot be completely folded after that point.  Therefore, only allowable folds
can be in the sequence.  It remains to show that performing an allowable fold
preserves foldability by a sequence of all-layers simple folds.  But performing
an allowable fold is equivalent to removing the smaller portion of paper to one
side of the fold.  Hence, it can only make more (allowable) folds possible, so
the mountain-valley pattern remains foldable.
\end{proof}

By Lemma \ref{allowable fold preserves foldability}, the problem of testing
foldability reduces to repeatedly finding allowable folds in the string.
Testing whether a fold at position $i$ is allowable can clearly be done in
$O(1+\min(i-1, N-i))$ time, by testing the boundary conditions and whether
$S[i-x] = \complement{S[i+x]}$ for $1 \leq x \leq \min(i-1, N-i)$.  Explicitly
testing all creases in this manner would yield an $O(n^2)$-time algorithm for
finding an allowable fold (if one exists).  Repeating this $O(n)$ times results
in a naive $O(n^3)$ algorithm for testing foldability.

This cubic bound can be improved by being a bit more careful. In $O(n^2)$ time,
we can determine for each crease $S[i]$ the largest value of $k$ for which
$S[i-x] = \complement{S[i+x]}$ for all $1 \leq x \leq k$.  \old{determine the
widest substring centered on it that obeys the complementarity condition (i.e.,
the widest substring centered on it that can be folded at that location).}
Using this information it is easy to test whether the fold at position $i$
is allowable.
\old{crease is allowable if $i-x$ or $i+x$ reaches an end of the string (i.e.,
$i-x=1$ or $i+x=N$).}
After making one of these allowable folds, we can in
$O(n)$ time update the value of $x$ for each crease, and hence maintain the
collection of allowable folds in linear time.  This gives an overall $O(n^2)$
bound, which we now proceed to improve further.

We present two efficient algorithms for folding strings.  The algorithm in
Section \ref{Suffix-Tree Algorithm} is based on suffix trees and runs in time
linear in the bit complexity of the input.  In Section \ref{Randomized
Algorithm}, we use randomization to obtain a simpler algorithm that runs in
$O(n)$ time.



\subsection{Suffix-Tree Algorithm}
\label{Suffix-Tree Algorithm}

In this section, we prove the following:

\begin{theorem} \label{1D all-layer suffix trees}
A string $S$ of length $N$ can be tested for all-layers simple foldability, in
time that is dominated by that to construct a suffix tree on $S$.
\end{theorem}

The difficulty with the time bound is that sorting the alphabet seems to be
required.  Other than the time to sort the alphabet, it is possible to
construct a suffix tree in $O(n)$ time \cite{Farach-1997}.  To sort the
alphabet in the comparison model, $O(n \log n')$ time suffices, where $n'$ is
the number of distinct input lengths.  In particular, if the input lengths are
encoded in binary, then the algorithm is linear in this bit complexity.  On a
RAM, the current state-of-the-art deterministic algorithm for integer sorting
\cite{Thorup-1998} uses $O(n (\log \log n)^2)$ time and linear space.

\medskip

\begin{proof}
Let $S^C$ be the complement string of $S$ (i.e., the complement of each letter
of $S$), and let $S^R$ be the reverse string of $S$.
The fold at position $i$ of $S$ is allowable precisely if
the first $\min(i-2,N-i+1)$ characters
of the suffix of $S^R$ starting in the $(N{-}i{+}2)$nd position are
identical to
the suffix of $S^C$ starting in the $(i{+}1)$st position,
and the single endpoint of $S$ ($S[1]$ if $i-1 <  N-i$, $S[N]$ if $N-i < i$)
has length less than or equal to its complement.

We build a single suffix tree containing all suffixes of $S^C$ and $S^R$ in
$O(n)$ time.  Further, we augment this tree with the capability to perform
least-common ancestor (LCA) queries in constant time after linear preprocessing
time \cite{Harel-Tarjan-1984,Schieber-Vishkin-1988}.  This LCA data structure
enables us to return the length of the longest prefix match of two given
suffixes in constant time.

To find the end-most possible fold, we can search for the
longest prefix match of $S^C [i+1]$ and $S^R [N-i+2]$, 
where the $j$th fold attempt takes place at $i = (j-1)/2$ if $j$ is odd,
and $i = N+1 - j/2$ if $j$ is even.
Thus the attempted folds alternate in from the left and right ends.
A fold can occur at $i$ if $S[i]$ equals $M$ or $V$, and
the length of the longest prefix match between $S^C [i+1]$ and $S^R [N-i+2]$
is $\min(i-1,N-i)$, or if the boundary condition above is satisfied.
We then perform this first legal fold, thus reducing the length of $S$.
We can continue our scan for the next fold by appropriately reducing
the length of the necessary longest prefix match to reflect the new
end of the string.  Note that the suffix tree remains unchanged,
and hence once one is computed, the folding process takes $O(n)$ time.
\end{proof}


\saurabh{The running time is dominated by time to create suffix tree which
        in turn dominates on time to sort the alphabet.  If we sort using
        randomization, we can sort in linear time, so the suffix tree algorithm
        can also serve the purpose of randomized linear time algorithm.  We
        should probably mention this.  The algorithm below is certainly much
        simpler and merits presence because of that.}
\saurabh{The above comment is not correct because it is not possible to
	sort in randomized linear time.}

\subsection{Randomized Algorithm}
\label{Randomized Algorithm}

In this section we describe a simple randomized algorithm that solves the 1D
all-layers simple-fold problem in $O(n)$ time.  There are two parts to the
algorithm:

  \begin{enumerate}
  \item assigning labels to the input lengths so that two lengths are 
        equal precisely if they have the same label;
        and
  \item finding and making allowable folds.
  \end{enumerate}

The first part is essentially element uniqueness, and can be solved in linear
expected time using hashing.  For example, the dynamic hashing method described
by Motwani and Raghavan \cite{Motwani-Raghavan-1995-dynamic-hashing} supports
insertions and existence queries in $O(1)$ expected time.  We can use this data
structure as follows.  For each input length, check whether it is already in
the hash table.  If it is not, we assign it a new unique identifier, and add it
to the hash table.  If it is, we use the existing unique identifier for that
value (stored in the hash table).  Let $n'$ denote the number of distinct
labels found in this process (or $2$, whichever is larger).

For the second part, we will show that each performed fold can be found in
$O(1+r)$ time, where $r$ is the number of creases removed by the discovered
fold (in other words, the minimum length to an end of the segment to be
folded).  However, it is possible that the algorithm makes a mistake, and that
some of the reported folds are not actually possible.  Fortunately, mistakes
can be detected quickly, and after $O(1)$ expected iterations the pattern will
be folded.  (Unless of course the pattern is not flat-foldable, in which case
the algorithm reports this fact correctly.)

The algorithm proceeds simultaneously from both ends of the segment, so that it
will find an allowable fold in time proportional to the minimum length from
either end.  At any point, the algorithm has a \emph{fingerprint} of the string
traversed before reaching that point, as well as a fingerprint of the
corresponding string immediately after that point (reversed and complemented).
These fingerprints are maintained in $O(1)$ time per step along the segment.
If the fingerprints match, then with high probability the underlying vectors
also match, and we have an allowable fold.  When we find such a fold (which
takes $O(1+r)$ time), the creases on the short side are discarded and the two
searches are restarted starting from both ends of the segment.
This process is repeated until no allowable folds are found, in which case
either the folding is complete (there are no creases left to perform) or the
crease pattern is not foldable by a sequence of all-layers simple folds (creases
remain).  In the former case, the folding sequence must be double checked
(again using $O(1+r)$ time per fold), and if it is incorrect, the entire
process is repeated with a new randomly chosen ``basis'' for fingerprints.

The fingerprints are based on Karp and Rabin's randomized string matching
algorithm \cite{Karp-Rabin-1987}.  We treat a substring as the base-$n'$
representation of an integer, where we use $0, \dots, n'-1$ to denote one of
the lengths, and $0$ or $1$ to denote a fold direction.  Then the fingerprint
of a substring is simply this integer modulo $p$ for a randomly chosen prime
$p$.  This fingerprint can be updated easily in constant time.  To add a symbol
to the end of the string, we multiply the current fingerprint by $n'$, and add
on the new symbol.  To add a symbol to the beginning of the string (which is
necessary for the reverse complement substring), we
add on the new symbol times
$(n')^k$ where $k$ is the current length of the string (we maintain $(n')^k
\bmod p$ throughout the computation).

Because an exact match is not required on the last length for a fold to be
allowable, both fingerprints on either side exclude the last symbol, and we
make a separate check that the length at the end is less than or equal to the
length onto which it folds.
Thus, given the appropriate fingerprints, we can check
whether a fold is allowable in $O(1)$ time.

By choosing the prime $p$ randomly from the range $2, \ldots, n^3$, the
probability that this algorithm makes a mistake after at most $n$ folds is
$O((\log n)/n)$; see \cite{Karp-Rabin-1987}.  More generally, if we choose $p$
from the range $2, \ldots, n^c$, then the probability of failure is $O(c (\log
n)/n^{c-2})$.  Thus, with high probability, the
algorithm gives a correct positive answer (it always gives correct negative
answers).  To obtain guaranteed correctness, we simply check the answer and
repeat the entire process upon failure.

\michael{For the final version, I want to check whether there is a better
reference than Karp and Rabin, something specifically on finding strings of the
form $u u w$.}


In conclusion, the algorithm we have presented proves the following result:

\begin{theorem} \label{1D all-layers randomized}
The 1D all-layers simple-fold problem can be solved in $O(n)$ time, both in
expectation and with high probability, on a machine supporting random numbers
and hashing of the input lengths.
\end{theorem}

\erik{
   OPEN 1: Is there an $\Omega(n log n)$ lower bound for the 1D all-layers
   problem with the fold directions and lengths specified as real numbers, in
   the comparison model?
 
     The obvious reduction would be from element distinctness, but set
     disjointness is another possibility.
 
 Assuming the answer to this question is YES, the 1D all-layers problem is
 finished.  (Unless we want to consider noncomparison models, but I'm not sure
 what you could gain from this.)}

\section{Orthogonal Simple Folds in 2D}
\label{ortho-2D}

%
%
\old{
(1) You use "V" to mean one thing in section 2 and another in section 5.


(2) Section 5 was the most interesting to me-- you should give
proofs for the "Note that" statements.
%
%
Where does "simple" enter into the argument?
%
%
Can you give a nonsimple example without an all-mountain or all-valley line?
%
%
}

\begin{figure}
\centerline{\psfig{file=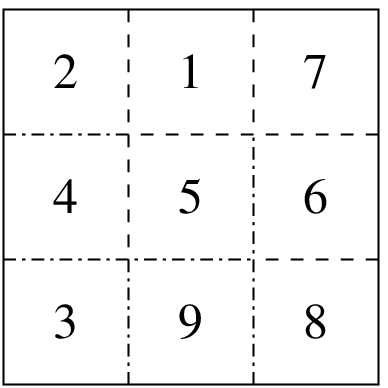}\hfil\psfig{file=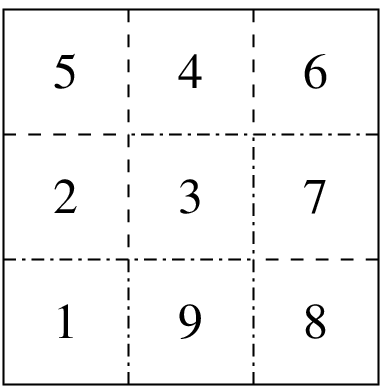}}
\caption{\label{hard map}
  Two maps that cannot be folded by simple folds, but can be folded flat.
  (These are challenging puzzles.)
  The numbering indicates the overlap order of faces.}
\end{figure}

In this section, we generalize our results for 1D simple folds to
\emph{orthogonal} 2D crease patterns, which consist only of horizontal and
vertical folds on a rectangular piece of paper, where horizontal and vertical
are defined by the sides of the rectangular paper.
In such a pattern, the creases must go all the way through the paper, because
every vertex of a flat-foldable crease pattern has degree at
least four \cite{Bern-Hayes-1996,Hull-1994}.
Hence, the crease pattern is a grid of creases (a \emph{map}),
although the space between grid lines may vary.
Edmonds \cite{Edmonds-1997-personal} observed that
orthogonal 2D mountain-valley patterns may be
flat-foldable but not by simple folds; see Figure \ref{hard map} for two
examples.
Recall from Section \ref{1D Some Layers} that the opposite holds in 1D:
one-layer and some-layers folds are equivalent to general flat-foldability.
In this section we simultaneously handle some-layers and all-layers
simple folds; with one-layer simple folds, only 1D maps are foldable.

To know what time bounds we desire, we must first discuss encoding the input.
A natural encoding of maps specifies the height of each row and the width of
each column, thereby using $n_1+n_2$ space for an $n_1 \times n_2$ grid.  The
mountain-valley assignment, however, requires $\Theta(n_1 n_2)$ space to
specify the direction for each edge of the grid.  Hence, our goal of linear
time amounts to being linear in $n = n_1 n_2$.

In a simply foldable mountain-valley pattern,
there must be at least one crease line, all the way across the
paper, that is entirely valley or mountain; otherwise,
the pattern would not permit any simple folds.
Furthermore, all such crease lines must be parallel;
otherwise, the vertex of intersection between two crossing crease lines would
not be locally flat-foldable.
Without loss of generality, assume that these crease lines are
horizontal, and let ${\cal H}$ denote the set of them.

We claim that all crease lines in ${\cal H}$ must be folded before any other
crease.
%
%
%
This is so because (1) folding along any vertical crease line $v$ will lead to
a mismatch of creases at the intersection of $v$ with any unfolded elements
of ${\cal H}$ and (2) horizontal crease lines not in ${\cal H}$ are not
entirely mountain or valley and hence cannot be folded before some vertical
fold is made.
Thus we have a corresponding 1D problem (with some- or all-layers folds)
to solve with the added necessary condition that the non-${\cal H}$
folds must match up appropriately after all the folds in ${\cal H}$ are made.
(The time spent checking this necessary condition can be attributed
to the non-${\cal H}$ folds that vanish after every fold.)
Because ${\cal H}$ contains at least one fold, performing the ${\cal H}$ folds
(strictly) reduces the size of the problem, and we continue.
The base case consists of just horizontal or vertical folds,
which corresponds to a 1D problem.  In summary we have

\begin{lemma} \label{order-is-immaterial}
If a crease pattern is foldable, it remains foldable after the folds
in ${\cal H}$ have been made in any feasible way considering ${\cal H}$
to be a 1D problem and ignoring other creases.
\end{lemma}

\old{
%
\begin{proof}
The paper is divided into rectangular columns by the folds in ${\cal V}$.  Each
column has some crease pattern on it.  The position and orientation (which side
up) of each column after all folds in ${\cal V}$ are made is fixed irrespective
of the order in which the folds in ${\cal V}$ are made.  This is so because if
we fix the leftmost end of the paper to be at origin, the columns will
alternate between top side up and bottom side up and between going left and
right.

Now because the crease pattern is foldable, the creases not in ${\cal V}$ match
for some order of folding ${\cal V}$.  And because in all orders of folding the
columns have the same position and orientation, the creases always match and
the composite crease pattern after all folds in ${\cal V}$ have been made is
the same irrespective of in which order the folds in ${\cal V}$ were made.  If
the paper is foldable this pattern must be foldable.  And hence after folding
${\cal V}$ in any feasible order considering it to be a 1D problem, the crease
pattern remains foldable.
\end{proof}
}

To find ${\cal H}$ quickly we maintain the number of mountain and valley
creases for each row and column of creases.  We maintain these numbers as
we make folds in ${\cal H}$.  To do this we traverse all the creases that
will vanish after a fold and decrement the corresponding numbers.  The cost
of this traversal is attributed to the vanishing creases.  Every time the
number of mountain or valley creases hits zero in a column or a row,
we add the row or column to a list to be used as the new ${\cal H}$ in
the next step.  Thus, we obtain

\begin{theorem}
Some-layers simple folding of an orthogonal crease pattern on a rectangular
piece of paper can be solved in deterministic linear time.  All-layers simple
folding in the same situation can be solved in randomized linear time, or
deterministic linear time plus the time required to sort the edge lengths.
\end{theorem}

This theorem easily generalizes to higher dimensions, with a running time
linear in $n = n_1 n_2 \cdots n_d$ plus possibly the time required to sort the
edge lengths.

\old{
%
\subsection{One- and some-layer folding}
For one- and some-layer folding we run the 1D one- and some-layer folding
algorithm on ${\cal V}$, with some extra work.  Every time we want to do a end-fold
we check if the crease pattern on the two sides of the fold are
anti-symmetric.  This can be done by walking on the two sides of the fold
in time linear in the number of creases that would get removed because of
the end-fold.  If we find a mismatch, that would mean that the input is
not foldable.  Similarly every time we want to do a crimp we check if the
crease pattern on the two sides of the two folds are anti-symmetric.  This
can be done by walking on the two sides of each fold in time linear in the
number of creases that would get removed because of the crimp.  Both while
doing the end-fold and crimp we need to maintain the number of mountain and
valley creases in each row and column.  This can be easily done by
decrementing these numbers each time we remove a crease.  Also note that
some columns would go out of existence.

Because the inductive step above uses time proportional to the number of
creases it removes, the running time of the algorithm is linear in the number
of creases in the input crease pattern.

\subsection{All layer folding}
For all-layers folding we run any of our 1D all-layers folding algorithm on
${\cal V}$, with some extra work.  Every time we want to do a fold, we check if the
crease pattern on the two sides of the fold is anti-symmetric.   Just as in
the one- and some-layers case this can be done in time linear in the number of
creases removed as a result of the fold.  As in one- and some-layers case we
maintain the mountain and valley crease numbers for each row and column.
Because everything else is done in time linear in the number of creases the
running time is determined by the 1D all-layers algorithm which is linear
deterministic in number of input bits or linear randomized.

\saurabh{The following is a nice observation about the beautiful structure
in 2D-ortho all-layers folding.  It could be useful.}

Note that in case of all-layers folding, the folding of all folds in ${\cal V}$
results into a single column whose width is equal to the widest column.
Hence if the input is foldable, the next step would deal with the pattern
on the widest column.  Thus we can solve this subproblem even before trying
to solve the 1D problem of folding ${\cal V}$.  We can keep doing this and we would
get a list of problems at each step all of which can be solved independently.

\saurabh{If we were to know that the input is foldable and is given to us
with some linear time processing.  Then we can find the folding in time
proportional to the number of folds which is better than time proportional
to the number of creases.  This is a bit fuzzy.  I would like to discuss
this with someone.  I am trying to get a "sort of" output sensitive algo.}

%
%
%



}

\section{Hardness of Simple Folds in 2D}
\label{Unrestricted 2D}

In this section we prove that the problem of deciding whether a 2D
axis-parallel mountain-valley pattern can be simply folded is (weakly) NP-hard,
if we allow the initial paper to be an arbitrary orthogonal polygon.  We also
show that it is (weakly) NP-hard to decide whether a mountain-valley pattern on
a square piece of paper can be folded by some-layers simple folds, if the
creases are allowed to be axis-parallel {\em plus} at a 45-degree angle.

Both hardness proofs are based on a reduction from an instance of {\sc
partition}, which is (weakly) NP-hard \cite{Garey-Johnson-1979}: given a set
$X$ of $n$ integers $a_1,a_2,\dots,a_n$ whose sum is $A$, does there exist a
set $S \subset X$ such that $\sum_{a\in S} a = {A/2}$?
For convenience we define the set $\bar S = X \setminus S$.
Also, without loss of generality, we assume that $a_1 \in S$.

\begin{figure}[htbp]
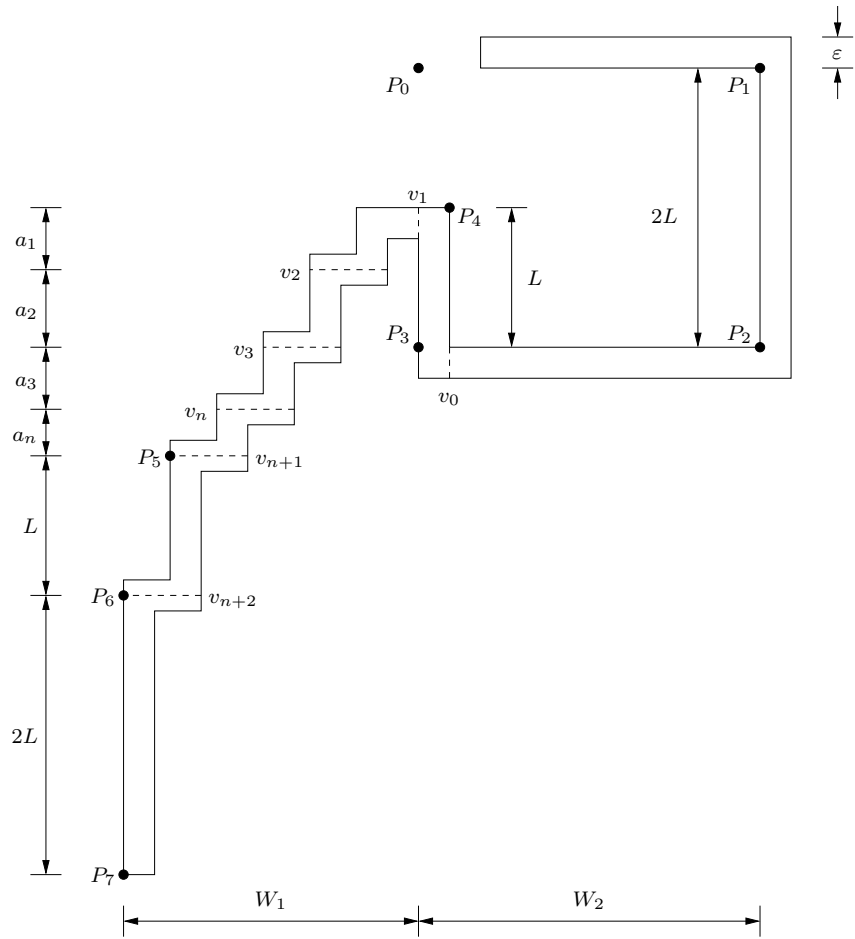

\centerline{\input 2dhard.pstex_t}
\caption{Hardness reduction from {\sc partition} problem.}
\label{fig-2d-hard}
\end{figure}

We transform an instance of the {\sc partition} problem into an
orthogonal 2D crease pattern on a orthogonal polygon, as shown in
Figure~\ref{fig-2d-hard}.
All creases are valleys.  There is a staircase of width
$\epsilon$, where $0 < \epsilon < 2/3$,
with one step of length $a_i$ corresponding to each element $a_i$ in $X$.
In addition, there are two final steps of length $L$ and $2L$,
where $L$ is chosen greater than $A/2$.
The total width $W_1$ of the staircase is chosen to be less then the width
$W_2$ of the frame attached to the staircase.

The main mechanism in the reduction is formed by the vertical creases $v_0$ and
$v_1$.  Basically, the first time we fold one of these two creases, the
staircase must fit within the frame, or else the second of these two creases is
blocked.  Then when we fold the other of these two creases, the staircase exits
the frame, enabling us to fold the remaining creases in the staircase.

\begin{lemma} \label{hardness-onlyif}
If the {\sc partition} instance has a solution, then the crease pattern in
Figure~\ref{fig-2d-hard} is simply foldable.
\end{lemma}

\begin{proof}
For $2 \leq i \leq n$, valley fold $v_i$ if 
exactly one of $a_{i-1}$ and $a_i$ is in $S$.
After these folds, as we travel along the steps corresponding to
$a_1, \dots, a_n$, we travel in the $-y$ direction for elements that
belong to $S$ and in the $+y$ direction for elements that belong
to $\bar S$.
Because the sums of elements of both $S$ and $\bar S$ are $A/2$, the point
$P_5$ has the same $y$-coordinate as the point $P_4$ after these folds.
Because $L > A/2$, the steps corresponding to $a_i$'s are confined to
remain in between the $y$ coordinates of points $P_1$ and $P_2$.
Because $P_5$ has the same $y$-coordinate as $P_4$ and because the vertical
distance between $P_5$ and $P_6$ is $L$, point $P_6$ will have the same
$y$-coordinate as either $P_1$ or $P_2$.

Now valley fold $v_{n+2}$.
Because the vertical distance between $P_6$ and $P_7$ is $2L$,
the $y$-coordinate
of $P_7$ will be same as that of $P_1$ or $P_2$ and the step between $P_6$
and $P_7$ will lie exactly between the $y$-coordinates of $P_1$ and $P_2$.
This situation is illustrated in Figure~\ref{fig-semifolded-staircase}.
\begin{figure}
\centerline{\input semifolded.pstex_t}
\caption{Semi-folded staircase confined between $y$ coordinates
        of $P_1$ and $P_2$.
  The top side of the paper is drawn white and the other side is drawn gray.}
\label{fig-semifolded-staircase}
\end{figure}

Now valley fold $v_1$.  Because $W_2 > W_1$, the partly folded
staircase, which currently lies between the $y$-coordinates of $P_1$
and $P_2$, fits within the rectangle $P_0 P_1 P_2 P_3$.
Now valley fold $v_0$.  We now have the semi-folded stairs on the right and
the rectangular frame $P_0 P_1 P_2 P_3$ on the left.
Finally, valley fold all of the remaining unfolded creases in the staircase.
This can be done because the rectangular frame is now on the left of $P_4$ and
all steps of the staircase are on the right of $P_4$.
\end{proof}

\begin{lemma} \label{hardness-if}
If the crease pattern in Figure~\ref{fig-2d-hard} is simply foldable, there
is a solution to the {\sc partition} instance.
\end{lemma}
 
\begin{proof}
If either $v_0$ or $v_1$ is folded without having the staircase
confined between the $y$-coordinates of $P_1$ and $P_2$, the rectangular
frame $P_0 P_1 P_2 P_3$ would intersect with the staircase and would make
the other of $v_0$ and $v_1$ impossible to fold.
Hence the staircase must be brought between the $y$-coordinates of
$P_1$ and $P_2$ before folding either $v_0$ or $v_1$.  Because the last
and the second-last steps of the staircase are of size $2L$ and $L$,
respectively, point $P_5$ must have the same coordinate as the point
$P_4$ when the staircase is confined between the $y$-coordinates of
$P_1$ and $P_2$.

As we travel from $P_4$ to $P_5$ along the staircase, we travel
equally in positive and negative $y$ directions along the steps
corresponding to the elements of $X$.  Hence the sum of elements along
whose steps we travel in negative $y$ direction is same as the sum of
elements along whose steps we travel in the $+y$ direction.  Thus
there is a solution to the {\sc partition} instance, if the crease
pattern in Figure~\ref{fig-2d-hard} is foldable.
\end{proof}

Lemmas~\ref{hardness-onlyif} and \ref{hardness-if} imply the following theorem.

\begin{theorem} \label{hardness-iff}
The problem of deciding simple foldability of an orthogonal piece of paper with
an orthogonal mountain-valley pattern is (weakly) NP-complete,
for all-layers, some-layers, and one-layer simple folds.
\end{theorem}

\ab{In the Appendix, we prove the following theorem, which shows that even}
\pa{Even}
on a rectangular piece of paper it is hard to decide foldability if,
besides axis-parallel, there are creases in diagonal
directions (45 degrees with respect to the axes):

\begin{theorem} \label{hardness-square}
It is (weakly) NP-complete to decide the simple foldability of an
(axis-parallel) square sheet of paper with a mountain-valley pattern having
axis-parallel creases and creases at the diagonal angles of 45 degrees with
respect to the axes, for both all-layers and some-layers simple folds.
\end{theorem}

\inappendix{
\begin{proof}
%
We transform an instance of the {\sc partition} problem to a 2D
crease pattern on an axis-parallel square having all creases either
orthogonal (axis-parallel) or at 45 degrees to the axes.

First, we establish a set of horizontal folds, evenly spaced and
alternating mountain and valley, which result in the paper becoming a
long thin rectangle (``strip''); these initial folds will be called
{\em I-folds}.

Now, a rectangular strip can be turned by 90 degrees by making a
single fold as shown in Figure~\ref{fig-turn-gadget}.  By making
several such turns we can get the strip into the shape of the initial
paper as in Figure~\ref{fig-2d-hard}, except that the corners at each
turn are ``shaved'' by 45-degree chamfers.  (A strip can readily be
folded in order to avoid these chamfers; however, it is easier to
describe our construction with the simpler single folds at each bend.)
Call these folds $d_1, d_2, d_3, \dots$.  Immediately after
each $d_i$ we make a fold parallel to the strip
to reduce its width.  See Figure~\ref{fig-dr-folds}.  Call these 
folds $r_1, r_2, r_3, \dots$ respectively.  Thus we make these folds
in the following order: $d_1, r_1, d_2, r_2, \dots$.
We refer to the $d_i$ folds as the {\em D-folds}, the $r_i$ folds as the
{\em R-folds}, and both kinds together as {\em DR-folds}.


\begin{figure}
\centerline{\psfig{file=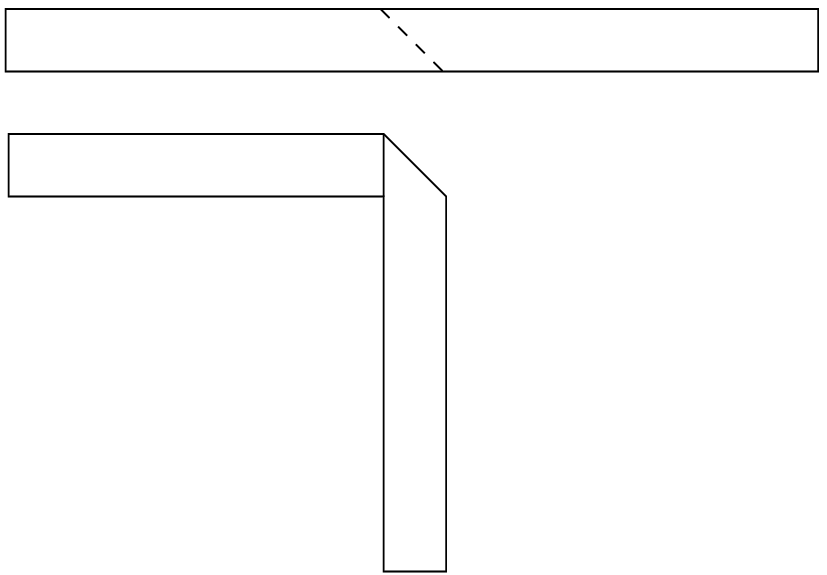,scale=0.5}}
\caption{Turning a strip.}
\label{fig-turn-gadget}
\end{figure}

Finally, we create valley folds, as shown in Figure~\ref{fig-2d-hard}.
We refer to these valleys as {\em F-folds}.  After making the F-folds,
we can unfold the paper to get the desired crease pattern.

\begin{figure}
\centerline{\psfig{file=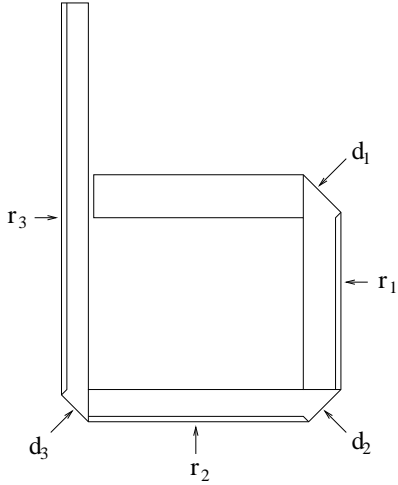,scale=0.45}}
\caption{Illustration of first few DR-folds used in construction.}
\label{fig-dr-folds}
\end{figure}

It is easy to see that if a solution to the {\sc partition} problem exists,
the above crease pattern can be folded by first making the I-folds,
followed by the DR-folds in the order in which they were created,
followed by the F-folds (as done in Figure~\ref{fig-2d-hard}).  We now
prove the other direction: if the crease pattern can be flat folded,
then the {\sc partition} problem has a solution.

Each D-fold intersects all I-folds.  And each R-fold intersects at least
one D-fold.  Hence none of the DR-folds can be made before all of the
(initial) I-folds are made.

Because $r_1$ intersects $d_1$ and was created after folding $d_1$, there
is a precedence constraint that in any valid folding $d_1$ occurs before
$r_1$.  Similarly $r_1$ occurs before $d_2$ and so on.  Thus the
DR-folds must occur in the order $d_1, r_1, d_2, r_2, \dots$.

Also none of the F-folds can be made before the corresponding R-folds which
it intersects are made.  Thus it is guaranteed that after I-folds
$d_1, r_1, d_2, r_2, r_3$ would be made in that order before any other folds.
This puts our rectangular frame $P_0, P_1, P_2, P_3$ in place as in
Figure~\ref{fig-2d-hard}.

Just as in proof of Lemma~\ref{hardness-if}, to enable folds $v_0$ and
$v_1$ to be made, the strip following $d_3$. must be folded so that it is
confined between the $y$-coordinates of $P_1$ and $P_2$.  For this proof,
this proof there is an additional constraint that no point on the strip
following $d_3$ should have an $x$-coordinate differing from that of
$P_0$ by more than $W_2$.  We can choose $W_2$ as small as we want, by
reducing the width of the strip that I-folds create.  In particular we
can choose $W_2$ to be smaller than all $a_i$'s.  Thus to meet the above
constraints all the lengths of the strip which correspond to $a_i$'s will
have to be vertical.  And just as in proof of Lemma~\ref{hardness-if},
there must exist a solution to {\sc partition} problem, if all these vertical
strips have to be between the $y$-coordinates of $P_1$ and $P_2$.
\end{proof}
}

The problem is open for the one-layer case.

\old{
%
\erik{
Nonorthogonal folds are the major open problem.  For all-layers folds, there is
an easy polynomial-time algorithm: check for folds of antisymmetry, apply one,
repeat.  Is there some kind of geometric generalization of suffix trees to
optimize this algorithm from $\Theta(n^2 \mathrm{polylog} n)$ down to $\Theta(n
\mathrm{polylog} n)$?  This is potentially very exciting!  For some-layers and
one-layer folds, any polynomial-time algorithm would be of interest.  Can the
ideas in Section \ref{1D Some Layers} be generalized?}

First, we consider the general case of crease patterns in 2D.
Following the discussion previously given after Lemma~\ref{allowable
fold preserves foldability}, we are able to determine foldability of a
crease $c$ by performing a linear-time search of the planar graph of
creases, on each side of $c$, in order to identify the largest
subgraph of ``matching'' (complementary) creases for $c$.  Crease $c$
is allowable, then, if after folding the paper along $c$, the matching
subgraph for $c$ lies completely within the polygonal region
corresponding to the set of points for which the fold resulted in an
increase in paper thickness.  This condition is readily checked in $O(n)$
time per crease (details omitted here), resulting in $O(n^2)$ time to
find a single allowable crease.  Applying this repeatedly gives
an $O(n^3)$ algorithm.  (Note that the
idea that improved the 1D problem from $O(n^3)$ to $O(n^2)$
\old{structure that allows an improvement to $O(n^2)$ in the 1D case}%
is not immediately applicable
in 2D, because the boundary of the paper can have complexity $\Omega(n)$
at any given stage in the folding, as opposed to the 1D case in which the
current shape of the ``paper'' is always a single line segment.)

Our goal is to look for analogously defined \emph{allowable folds}, which can
be thought of as lines of antisymmetry.  That is, folds on one side of the line
must have the opposite direction of the symmetric folds on the other side of
the line.  Furthermore, every crease along an allowable crease line must have the
same direction.  Again, performing an allowable fold corresponds to chopping
off a portion of the rectangle to be folded, and thus certainly makes progress.
}

\old{
** Commenting out for now, until we get more interesting results...

\section{Hardness Results}
\label{Hardness Results}

While we have given efficient algorithms for determining foldability
in our models of simple folding, we now show that some versions of our
problem become hard when an optimization criterion is imposed.
Specifically, if our goal is to minimize the total size (length) of
the flat origami in 1D, and we are allowed at each crease to fold (in
either direction) or not to fold at all, then the problem of deciding
whether a given size is achievable is NP-hard.  Our reduction is from an
instance of {\sc Partition}: Given integers $a_1,a_2,\dots,a_n$ whose sum
is $A$, does there exist a subset $S\subset\{1,2,...,n\}$ such that
$\sum_{i\in S}a_i={A\over 2}$?

The main ``gadget'' in our hardness proof is similar to one described
in \cite{Arkin-Fekete-Mitchell-Skiena-1999-manuscript}. Given an
instance of {\sc Partition}, we construct the following
crease pattern: $A\,c\,{A\over 2}\,c\,a_1\,c\,a_2\,c\,\cdots\,
a_n\,c\,{A\over 2}\,c\,A$. Each $c$ is a potential crease which can either be
folded (in either direction) or left straight (unfolded).  The proof
of the following lemma follows almost directly from
\cite{Arkin-Fekete-Mitchell-Skiena-1999-manuscript}:

\begin{lemma}
A folding of the crease pattern $Ac{A\over 2}ca_1ca_2c
\cdots a_nc{A\over 2}cA$ such that the length of the resulting segment
is exactly $A$ is possible if and only if the {\sc Partition} instance
is a ``yes'' instance.
\end{lemma}

The following theorem refers to the some-layers model, which we
have seen is equivalent to the one-layer model.
Its proof is immediate from the lemma above.

\begin{theorem} 
The following problem is (weakly) NP-hard in the some-layers model
of folding: 
Given a crease pattern (each crease
of which can be mountain, valley, or not folded at all), fold the
segment so as to minimize the length of the folded object.
\end{theorem}


\estie{I don't believe the following proof, which is my interpretation
of Erik's picture! Well actually part (b) is just a special case of
the lemma above but part (a) I think is wrong. The alleged opt proof
folds the entire spirals and some of the subset sum guys. A better
folding will fold the entire spirals except for the innermost one,
and then all the subset sum, which then won't be equal to anything,
but who cares when all we care about is the max number of folds not
length. I don't see a way to fix this.}

\joe{assuming others agree, I propose to comment out the
rest of this section

\begin{theorem} 
The following two problems are (weakly) NP-hard: Given a crease
pattern and a {\em partial} mountain-valley pattern, fold the segment
so as either (a) to maximize the number of folds made, or (b) to minimize
the length of the folded object.  
\end{theorem}

\begin{proof} 
We append to the construction of the segment given above on the left
``many'' (such as $5n$) intervals of length $A$, and ``many''
intervals of length $A$ on the right. All the new creases on the left
are valleys, and the new creases on the right are mountains.  This
yields the segment $AVAVA\cdots Ac{A\over 2}ca_1ca_2c \cdots
a_nc{A\over 2}cAMAM\cdots A$. Note that the left intervals of length $A$ form
an inward spiral, and the right intervals of length $A$ for an outward
spiral. The only way both spirals can be folded is to have the inner
part of the segment, $Ac{A\over 2}ca_1ca_2c \cdots a_nc{A\over 2}cA$,
folded to length exactly~$A$.
\end{proof}
}

\old{Erik's original picture caption:(can't see/read in his
eps file, as it gets clipped)

Here is the folding I have in mind.  It is realizable precisely if
there is a subset B of the $a_i$'s such that exactly one of $a_1$ and $a_m$
are in B, and the sum of the values in B equals the sum of the values
not in B.  Furthermore, if we make the number of folds at the
extremities extremely large (e.g., 5n), then I claim this folding (if
realizable) maximizes the number of performed folds and minimizes the
length of the folded object.
}

\joe{I suggest keeping the open problem list below ``confidential''
for now and putting it inside a comment statement.}

\erik{Agreed.}

\comment{
The complexity of several problems remains open even in the 1D case:

Given a (complete) mountain-valley pattern, fold the segment in the
one-layer model so as to either (a) maximize the number of folds made,
or (b) minimize the length of the folded object.

Given a mountain-valley pattern, minimize the number of fold reversals
to make it flat-foldable in the some-layers model.

Are some of these problems hard for the all-layers model?
Or can they be solved in polynomial time?

Note that the all-layers model can no longer be solved by any of our
algorithms in Section~\ref{1D All Layers}, which relied on the fact
that making any feasible fold is ``progress'' which cannot hurt us
later on. This is no longer true for the optimization problems here.

Here are some old interesting questions which we believe are NP-hard.

Consider an unfolded piece of paper (such as an open map) that have been
mangled somewhat in an unsuccessful attempt to fold it.  As a result, some of
the creases are either \emph{ambiguous} (we cannot tell whether they are
mountains or valleys) or \emph{incorrect} (they appear to be one direction,
but folding requires them to be the other direction).
This motivates the following problems:

\problem Ambiguous Simple Folding:
Given a crease pattern and a partial mountain-valley assignment,
is there a simple folding that satisfies the specified mountains and valleys?

\problem Optimal Simple Folding:
Given a crease pattern and a mountain-valley assignment,
find a simple folding that minimizes the number of mountain-valley violations
(if one exists).

}

}

\section{No Mountain-Valley Assignments}

An interesting case to consider is when the creases do not have
mountain-valley assignment:  any crease can be folded in either
direction.  Even with this flexibility, we are able to show that
the problem is hard\ab{ (see the Appendix for the proof)}:

\begin{theorem}
\label{thm:no-mountain-valley}
The problem of deciding the simple foldability of an orthogonal piece of paper
with a crease pattern (without a mountain-valley assignment) is (weakly)
NP-complete, for both all-layers and some-layers simple folds.
\end{theorem}

\inappendix{
\begin{proof}
For the all-layers case, the proof of Theorem~\ref{hardness-iff} works
without mountain-valley assignments as well.  This is so because
the staircase must be confined as before to make both turns $v_1$
and $v_2$ in either direction.  If the staircase is not confined
before either of $v_1$ or $v_2$ is made in either direction, it
will overlap with the frame, and, in the all-layers case, as soon as
two layers of paper overlap they are ``stuck'' together.

\begin{figure}
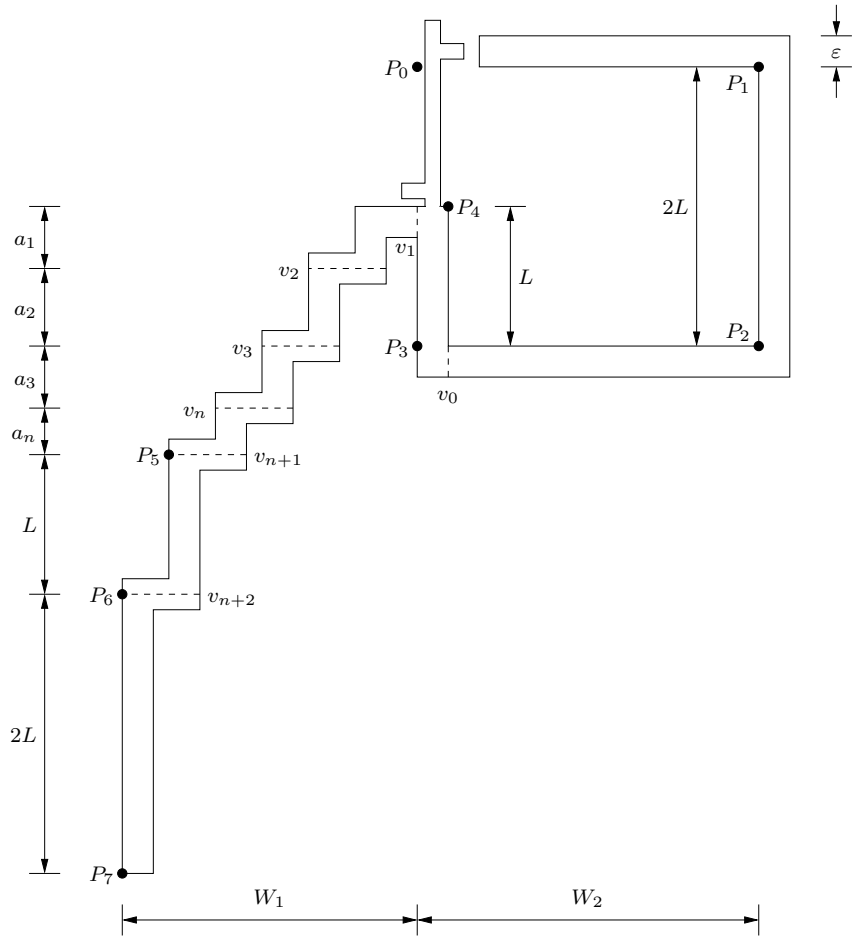

\centerline{\input noassign.pstex_t}
\caption{Hardness reduction when no mountain-valley assignment is given.}
\label{fig-noassign}
\end{figure}

\saurabh{I am sure the proof can be modified to just have 45 degree
directions apart from horizontal and vertical.  But I hope it is
not important to do it.}

For the some-layers case the proof of Theorem~\ref{hardness-iff} does
not work, as the folds $v_0$ and $v_1$ can be made in opposite
directions, and so a folding exists whether or not a partition
exists. We modify the construction to ensure that $v_0$ and $v_1$ must
be folded in the same direction.  See Figure~\ref{fig-noassign}, and
the more detailed Figure~\ref{fig-zoom}.  There are only two
differences between this construction and the one in
Figure~\ref{fig-2d-hard}.  First is the extra piece of paper ({\em
flap}) attached at the top of the staircase.  Second is the addition
of the fold of the flap, and three ``crimps'' shown in
Figure~\ref{fig-zoom}. When creating the crease pattern, these new
folds are made before the folds $v_0$ and $v_1$.  Each crimp consists of two
folds very close to each other, changing the shape of our
construction only infinitesimally.

It is easy to argue that if there is a solution to the {\sc partition}
problem, then our construction can be folded.  This can be done by
first folding the flap, followed by crimp $c_0$, followed by crimps
$c_1$, $c_2$ and then following the algorithm described in proof of
Lemma~\ref{hardness-onlyif}.

\begin{figure}
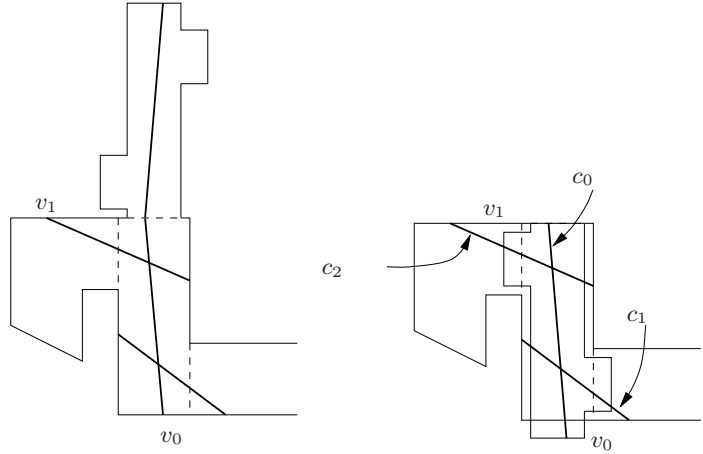

\centerline{\input zoom.pstex_t}
\caption{Interesting part of construction for hardness reduction.}
\label{fig-zoom}
\end{figure}

We now prove the other direction; that is, if our construction is
foldable, then there is a solution to the {\sc partition} problem.
We start by noting the following: Given a crease pattern in which two
folds intersect at an angle other than 90 degrees, it is easy to tell
which of the two folds must be folded first in any legal folding. This
is because the second fold must be a mirror image through the first
fold. If the angle of intersection is not 90 degrees, then the second
fold does not form a straight line in the crease pattern, but rather
is two line segments reflected around the first fold. (If two creases
meet at a 90-degree angle, and no mountain-valley assignment is given,
then there are two possible orders of folding the two creases.)  As a
consequence, while constructing a crease pattern, if a crimp or a fold
$c_y$ is folded after another crimp or fold $c_x$ and if $c_y$
intersects $c_x$ at any angle other than $90$ degrees, then $c_x$ must
be folded before $c_y$ can be folded in any legal folding of this
crease pattern.

Figure~\ref{fig-crimp-cross} illustrates two crimps intersecting at $45$
degrees and the crease pattern they create.

\begin{figure}
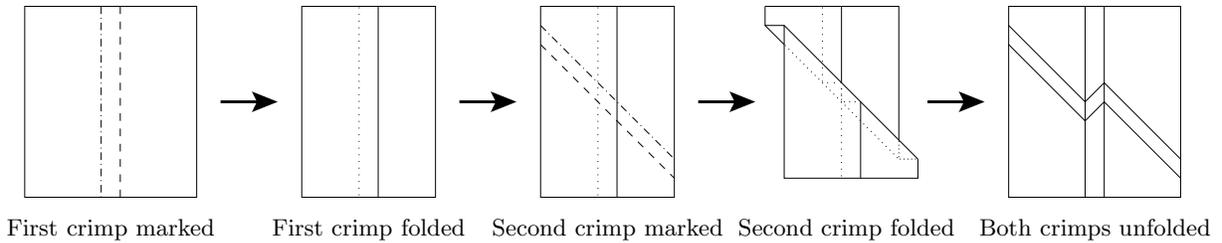

\centerline{\input crimp_crossing.pstex_t}
\caption{Crimps intersecting at an angle other than $90$ degrees cannot be
folded out of order.}
\label{fig-crimp-cross}
\end{figure}

In our construction, from the above discussion, the flap must be
folded before crimp $c_0$ can be folded, which in turn needs to be
folded before crimps $c_1$ and $c_2$ can be folded.  Further, crimps
$c_1$ and $c_2$ need to be folded before folds $v_0$ and $v_1$ can
respectively be folded.  Thus, before either $v_0$ or $v_1$ can be
folded, the flap must be folded.  Once the flap is folded in either
direction, $v_0$ and $v_1$ are forced to fold in the same direction.
With this constraint, the rest of the proof is the same as that of
Lemma~\ref{hardness-if}.
\end{proof}
}

The problem is open for the one-layer case.

\old{

Here we prove Theorem~\ref{hardness-square}.

As in section~\ref{Unrestricted 2D} we reduce from {\sc partition}.

We transform an instance of the {\sc partition} problem to a 2D
crease pattern on an axis-parallel square having all creases either
orthogonal (axis-parallel) or at 45 degrees to the axes.

First, we establish a set of horizontal folds, evenly spaced and
alternating mountain and valley, which result in the paper becoming a
long thin rectangle (``strip''); these initial folds will be called
{\em I-folds}.

Now, a rectangular strip can be turned by 90 degrees by making a
single fold as shown in Figure~\ref{fig-turn-gadget}.  By making
several such turns we can get the strip into the shape of the initial
paper as in Figure~\ref{fig-2d-hard}, except that the corners at each
turn are ``shaved'' by 45-degree chamfers.  (A strip can readily be
folded in order to avoid these chamfers; however, it is easier to
describe our construction with the simpler single folds at each bend.)
We refer to these folds as {\em D-folds}.

\begin{figure}
\centerline{\psfig{file=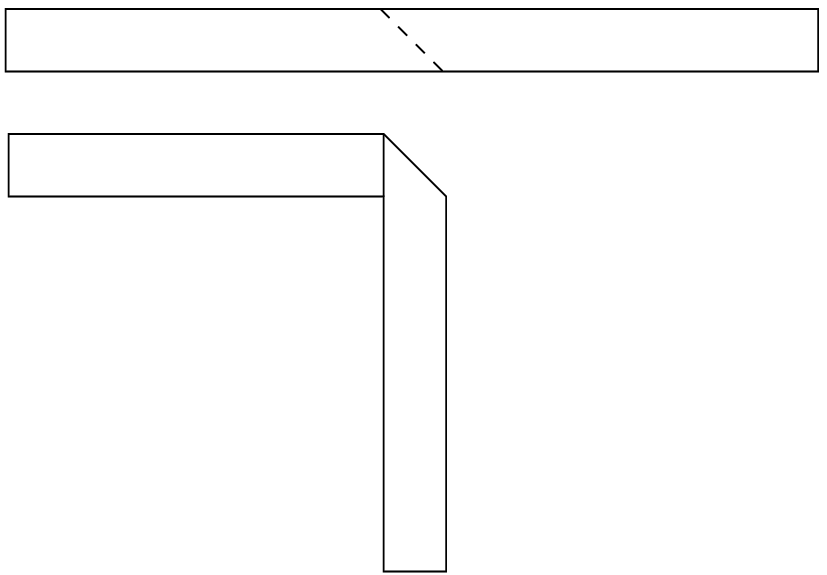,height=1.0in}}
\caption{Turning a strip.}
\label{fig-turn-gadget}
\end{figure}

By a {\em crimp} we will mean pair of parallel and nearly adjacent
mountain and valley folds.  Once the strip is folded into the
desired orthogonal shape, we start from one end of the strip and make
crimps along the direction parallel to the strip.  Each crimp ends at
a turn.  From a turn in the strip we start a new crimp orthogonal to
and intersecting the previous one.  We keep doing this until we reach
the end of the strip.  At the end of the strip we make a crimp in the
direction orthogonal to the strip and then another crimp in the
direction parallel to the strip of paper.  We now continue to make
crimps and traverse the strip in the opposite direction.  This
construction is shown in Figure~\ref{fig-square-hard}.  We refer to
the set of crimps made during the first traversal of the strip from
end to end as {\em C1-folds} and those made during the reverse
traversal as {\em C2-folds}.  All together we refer to the crimps as
{\em C-folds}.

Finally, we create valley folds, as shown in Figure~\ref{fig-2d-hard}.
We refer to these valleys as {\em F-folds}.  After making the F-folds,
we can unfold the paper to get the desired crease pattern.

\begin{figure}
\centerline{\input squarehard.pstex_t}
\caption{Illustration of the set of crimp folds used in the construction.}
\label{fig-square-hard}
\end{figure}

It is easy to see that if a solution to the partition problem exists,
the above crease pattern can be folded by first making the I-folds,
followed by the D-folds, followed by all the C-folds in order,
followed by the F-folds (as done in Figure~\ref{fig-2d-hard}).  We now
prove the other direction: if the crease pattern can be flat folded,
then the partition problem has a solution.

Each D-fold intersects all I-folds.  Hence none of the
D-folds can be made before all of the (initial) I-folds are made.

Each D-fold is intersected by four crimps.  None of these four
crimps can be made before the D-fold is made.

Also note that if a crimp $C_2$ is made immediately after another
crimp $C_1$, in our construction, then in any simple folding crimp
$C_1$ must be folded before crimp $C_2$.  Hence, there is a total
order amongst crimps and the order in which they can be made is the
same as the order in which they were made in our construction.

Any F-fold crosses two crimps (one from each set).  These crimps must
therefore be folded before the F-fold can be made.  Hence, before any
F-fold is made, all the crimps in C1-folds must have been folded.

Because a crimp from C1-folds intersects each D-fold, when all the C1-fold
crimps have been folded, all the D-folds must have also been made.

Thus, our strip must get into the shape of Figure~\ref{fig-2d-hard}
{\em before} any F-fold is made.  But once the paper is in that shape
with all of the F-folds and C2-folds still to be made, the F-folds can
be made only if there exists a solution to the partition problem; this
we have seen in Section~\ref{Unrestricted 2D}.  The C2-folds do not
make a difference.  They can be made anytime when feasible; they only
shrink the width of the staircase by a very small amount.
}

\section{Conclusion}

We have presented efficient algorithms for deciding flat foldability of a map
(rectangle with horizontal and vertical creases) via a sequence of simple
folds, for any of three different restrictions on the number of layers that can
be folded at once.  In the all-layers model, there may be several solution
sequences of simple folds, and they can vary significantly in length; for
example, in a 1D pattern that alternates mountain and valley, there is a
sequence with roughly $\log n$ folds and a sequence with roughly $n$ folds.
(In contrast, the shape of the final folded state is independent of the folding
process, depending only in the crease pattern.)
Jeff Erickson\footnote{Personal communication, March 2001.} observed that there
is also a polynomial-time algorithm for minimizing the length of the
simple-fold sequence: the one-dimensional subproblems are forced, and in each
one-dimensional subproblem, we can use dynamic programming on the $O(n^2)$
substrings of the mountain-valley pattern.
This optimization problem is of course trivial
for one-layer simple folds (the number of folds equals the number of creases),
but it remains open for some-layers simple folds, where there is an interesting
interplay between the efficiency of all-layers folds and the power of
one-layer folds.

On the complexity side, we have shown that slight generalizations of the basic
map-folding problem are (weakly) NP-complete.
However, there still remains a gap.  For example, what is the complexity of
deciding simple foldability of an orthogonal crease pattern on an orthogonally
convex piece of paper?  Even more simply, what is the complexity of deciding
simple foldability of an orthogonal crease pattern on a convex piece of paper,
or even a non-axis-aligned rectangle?
These variations possess the needed difficulty that making one fold may produce
a piece of paper that is no longer a subset of the original piece of paper;
thus, it is not clear that making a fold always makes progress.
Another direction to consider is nonrectangular maps, for example,
a triangular map whose faces are unit equilateral triangles
(polyiamonds).  We conjecture that deciding simple foldability is again
(weakly) NP-complete in this context.
Also, for the problems that we show to be weakly NP-hard,
it remains open whether there
are pseudopolynomial-time algorithms for solving simple foldability,
or whether the problems are strongly NP-complete.


Our study of special cases of crease patterns may also be interesting in the
context of general flat foldings.  Here the goal would be to strengthen Bern
and Hayes's NP-hardness result \cite{Bern-Hayes-1996} to special crease
patterns, or perhaps more interesting, to find special cases in which flat
foldability is polynomially testable.  One special case of particular interest,
posed by Edmonds \cite{Edmonds-1997-personal}, is an $m \times n$ grid
with a prescribed mountain-valley assignment.
Along these lines, Justin \cite{Justin-1994} observed that even $2 \times n$
maps in which every $2 \times 2$ submap is flat-foldable may not be totally
flat-foldable.
Di Francesco \cite{DiFrancesco-2000} suggests that a useful algebraic structure
in $1 \times n$ map folding may generalize.
However, we do not even know whether testing general flat foldability is
NP-hard for the cases in which testing simple foldability is NP-hard:
orthogonal polygons with orthogonal creases, and rectangles with orthogonal and
$45^\circ$ creases.  Our hardness reductions rely on the restriction to simple
folds.

\section*{Acknowledgments}
We thank Jack Edmonds for helpful discussions which inspired this research,
in particular for posing some of the problems addressed here.
We also thank Joseph O'Rourke and an anonymous referee for extensive comments
which greatly improved this paper.
E.~Arkin and J.~Mitchell thank Mike Todd for posing algorithmic versions
of the map-folder's problem in a geometry seminar at Cornell.

E. Arkin acknowledges support from the National Science Foundation
(CCR-9732221) and HRL Laboratories. M.~Bender acknowledges support
from HRL Laboratories.  J.~Mitchell acknowledges support from HRL
Laboratories, the National Science Foundation (CCR-9732221), NASA Ames
Research Center, Northrop-Grumman Corporation, Sandia National Labs,
Seagull Technology, and Sun Microsystems.

\section*{Addendum}
We recently learned that Calinescu, Karloff, and Thorup
\cite{Calinescu-Karloff-Thorup-2000} independently discovered linear-time
algorithms for some-layers simple foldability in the 1D and 2D orthogonal
cases.

\bibliography{algs,biology,complexity,linkage,origami,random,robotics,sorting,strings,unfolding}
\bibliographystyle{plain}

\end{document}

%% file: various_layers_1d.pstex_t
\begin{picture}(0,0)%
\includegraphics{various_layers_1d.pstex}%
\end{picture}%
\setlength{\unitlength}{3236sp}%
\begingroup\makeatletter\ifx\SetFigFont\undefined%
\gdef\SetFigFont#1#2#3#4#5{%
  \reset@font\fontsize{#1}{#2pt}%
  \fontfamily{#3}\fontseries{#4}\fontshape{#5}%
  \selectfont}%
\fi\endgroup%
\begin{picture}(9149,2112)(1514,-2161)
\put(3001,-436){\makebox(0,0)[b]{\smash{{\SetFigFont{10}{12.0}{\rmdefault}{\mddefault}{\updefault}{\color[rgb]{0,0,0}Starting configuration}%
}}}}
\put(3001,-661){\makebox(0,0)[b]{\smash{{\SetFigFont{10}{12.0}{\rmdefault}{\mddefault}{\updefault}{\color[rgb]{0,0,0}for a simple fold}%
}}}}
\put(3001,-1186){\makebox(0,0)[b]{\smash{{\SetFigFont{9}{10.8}{\rmdefault}{\mddefault}{\updefault}{\color[rgb]{0,0,0}top side}%
}}}}
\put(2776,-1486){\makebox(0,0)[rb]{\smash{{\SetFigFont{9}{10.8}{\rmdefault}{\mddefault}{\updefault}{\color[rgb]{0,0,0}valley fold left side}%
}}}}
\put(3001,-2161){\makebox(0,0)[b]{\smash{{\SetFigFont{9}{10.8}{\rmdefault}{\mddefault}{\updefault}{\color[rgb]{0,0,0}crease location}%
}}}}
\put(5401,-661){\makebox(0,0)[b]{\smash{{\SetFigFont{10}{12.0}{\rmdefault}{\mddefault}{\updefault}{\color[rgb]{0,0,0}(one-layer model)}%
}}}}
\put(5401,-436){\makebox(0,0)[b]{\smash{{\SetFigFont{10}{12.0}{\rmdefault}{\mddefault}{\updefault}{\color[rgb]{0,0,0}1-layer simple fold}%
}}}}
\put(9601,-436){\makebox(0,0)[b]{\smash{{\SetFigFont{10}{12.0}{\rmdefault}{\mddefault}{\updefault}{\color[rgb]{0,0,0}3-layer simple fold}%
}}}}
\put(9601,-661){\makebox(0,0)[b]{\smash{{\SetFigFont{10}{12.0}{\rmdefault}{\mddefault}{\updefault}{\color[rgb]{0,0,0}(all-layers model)}%
}}}}
\put(7501,-436){\makebox(0,0)[b]{\smash{{\SetFigFont{10}{12.0}{\rmdefault}{\mddefault}{\updefault}{\color[rgb]{0,0,0}2-layer simple fold}%
}}}}
\put(7501,-661){\makebox(0,0)[b]{\smash{{\SetFigFont{10}{12.0}{\rmdefault}{\mddefault}{\updefault}{\color[rgb]{0,0,0}(some-layers model)}%
}}}}
\end{picture}%

%% file: various_layers_2d.pstex_t
\begin{picture}(0,0)%
\includegraphics{various_layers_2d.pstex}%
\end{picture}%
\setlength{\unitlength}{3947sp}%
\begingroup\makeatletter\ifx\SetFigFont\undefined%
\gdef\SetFigFont#1#2#3#4#5{%
  \reset@font\fontsize{#1}{#2pt}%
  \fontfamily{#3}\fontseries{#4}\fontshape{#5}%
  \selectfont}%
\fi\endgroup%
\begin{picture}(4630,1524)(4083,-2773)
\put(4576,-2588){\makebox(0,0)[rb]{\smash{{\SetFigFont{10}{12.0}{\rmdefault}{\mddefault}{\updefault}{\color[rgb]{0,0,0}2 layers}%
}}}}
\put(4576,-1838){\makebox(0,0)[rb]{\smash{{\SetFigFont{10}{12.0}{\rmdefault}{\mddefault}{\updefault}{\color[rgb]{0,0,0}1 layer }%
}}}}
\put(4576,-2288){\makebox(0,0)[rb]{\smash{{\SetFigFont{10}{12.0}{\rmdefault}{\mddefault}{\updefault}{\color[rgb]{0,0,0}3 layers}%
}}}}
\end{picture}%

%% file: model_power.pstex_t
\begin{picture}(0,0)%
\includegraphics{model_power.pstex}%
\end{picture}%
\setlength{\unitlength}{3947sp}%
\begingroup\makeatletter\ifx\SetFigFont\undefined%
\gdef\SetFigFont#1#2#3#4#5{%
  \reset@font\fontsize{#1}{#2pt}%
  \fontfamily{#3}\fontseries{#4}\fontshape{#5}%
  \selectfont}%
\fi\endgroup%
\begin{picture}(6981,2332)(1189,-2681)
\put(4801,-2251){\makebox(0,0)[lb]{\smash{{\SetFigFont{9}{10.8}{\rmdefault}{\mddefault}{\updefault}{\color[rgb]{0,0,0}but not flat-foldable by}%
}}}}
\put(1201,-2446){\makebox(0,0)[lb]{\smash{{\SetFigFont{9}{10.8}{\rmdefault}{\mddefault}{\updefault}{\color[rgb]{0,0,0}all-layers simple folds}%
}}}}
\put(1201,-2251){\makebox(0,0)[lb]{\smash{{\SetFigFont{9}{10.8}{\rmdefault}{\mddefault}{\updefault}{\color[rgb]{0,0,0}but not flat-foldable by}%
}}}}
\put(1201,-2056){\makebox(0,0)[lb]{\smash{{\SetFigFont{9}{10.8}{\rmdefault}{\mddefault}{\updefault}{\color[rgb]{0,0,0}one-layer simple folds}%
}}}}
\put(1201,-1861){\makebox(0,0)[lb]{\smash{{\SetFigFont{9}{10.8}{\rmdefault}{\mddefault}{\updefault}{\color[rgb]{0,0,0}flat-foldable by}%
}}}}
\put(3001,-1861){\makebox(0,0)[lb]{\smash{{\SetFigFont{9}{10.8}{\rmdefault}{\mddefault}{\updefault}{\color[rgb]{0,0,0}flat-foldable by}%
}}}}
\put(3001,-2056){\makebox(0,0)[lb]{\smash{{\SetFigFont{9}{10.8}{\rmdefault}{\mddefault}{\updefault}{\color[rgb]{0,0,0}all-layers simple folds}%
}}}}
\put(3001,-2251){\makebox(0,0)[lb]{\smash{{\SetFigFont{9}{10.8}{\rmdefault}{\mddefault}{\updefault}{\color[rgb]{0,0,0}but not flat-foldable by}%
}}}}
\put(3001,-2446){\makebox(0,0)[lb]{\smash{{\SetFigFont{9}{10.8}{\rmdefault}{\mddefault}{\updefault}{\color[rgb]{0,0,0}one-layer simple folds}%
}}}}
\put(4801,-2641){\makebox(0,0)[lb]{\smash{{\SetFigFont{9}{10.8}{\rmdefault}{\mddefault}{\updefault}{\color[rgb]{0,0,0}all-layers simple folds}%
}}}}
\put(4801,-2446){\makebox(0,0)[lb]{\smash{{\SetFigFont{9}{10.8}{\rmdefault}{\mddefault}{\updefault}{\color[rgb]{0,0,0}one-layer and/or }%
}}}}
\put(4801,-2056){\makebox(0,0)[lb]{\smash{{\SetFigFont{9}{10.8}{\rmdefault}{\mddefault}{\updefault}{\color[rgb]{0,0,0}some-layers simple folds}%
}}}}
\put(4801,-1861){\makebox(0,0)[lb]{\smash{{\SetFigFont{9}{10.8}{\rmdefault}{\mddefault}{\updefault}{\color[rgb]{0,0,0}flat-foldable by}%
}}}}
\put(6601,-1861){\makebox(0,0)[lb]{\smash{{\SetFigFont{9}{10.8}{\rmdefault}{\mddefault}{\updefault}{\color[rgb]{0,0,0}flat-foldable by}%
}}}}
\put(6601,-2056){\makebox(0,0)[lb]{\smash{{\SetFigFont{9}{10.8}{\rmdefault}{\mddefault}{\updefault}{\color[rgb]{0,0,0}general origami folding}%
}}}}
\put(6601,-2251){\makebox(0,0)[lb]{\smash{{\SetFigFont{9}{10.8}{\rmdefault}{\mddefault}{\updefault}{\color[rgb]{0,0,0}but not flat-foldable by}%
}}}}
\put(6601,-2446){\makebox(0,0)[lb]{\smash{{\SetFigFont{9}{10.8}{\rmdefault}{\mddefault}{\updefault}{\color[rgb]{0,0,0}any simple folds}%
}}}}
\end{picture}%

%% file: local_ops.pstex_t
\begin{picture}(0,0)%
\epsfig{file=local_ops.pstex}%
\end{picture}%
\setlength{\unitlength}{3552sp}%
\begingroup\makeatletter\ifx\SetFigFont\undefined%
\gdef\SetFigFont#1#2#3#4#5{%
  \reset@font\fontsize{#1}{#2pt}%
  \fontfamily{#3}\fontseries{#4}\fontshape{#5}%
  \selectfont}%
\fi\endgroup%
\begin{picture}(6685,1644)(889,-2236)
\put(2101,-811){\makebox(0,0)[b]{\smash{\SetFigFont{11}{13.2}{\rmdefault}{\mddefault}{\updefault}M}}}
\put(2701,-811){\makebox(0,0)[b]{\smash{\SetFigFont{11}{13.2}{\rmdefault}{\mddefault}{\updefault}V}}}
\put(2101,-1186){\makebox(0,0)[b]{\smash{\SetFigFont{11}{13.2}{\rmdefault}{\mddefault}{\updefault}$c_i$}}}
\put(4501,-736){\makebox(0,0)[b]{\smash{\SetFigFont{11}{13.2}{\rmdefault}{\mddefault}{\updefault}Crimp}}}
\put(4501,-1636){\makebox(0,0)[b]{\smash{\SetFigFont{11}{13.2}{\rmdefault}{\mddefault}{\updefault}End fold}}}
\put(2701,-1186){\makebox(0,0)[b]{\smash{\SetFigFont{11}{13.2}{\rmdefault}{\mddefault}{\updefault}$c_{i+1}$}}}
\put(6301,-811){\makebox(0,0)[b]{\smash{\SetFigFont{11}{13.2}{\rmdefault}{\mddefault}{\updefault}$c_i$}}}
\put(5701,-1336){\makebox(0,0)[b]{\smash{\SetFigFont{11}{13.2}{\rmdefault}{\mddefault}{\updefault}$c_{i+1}$}}}
\put(3151,-2086){\makebox(0,0)[b]{\smash{\SetFigFont{11}{13.2}{\rmdefault}{\mddefault}{\updefault}$c_n$}}}
\put(3151,-1711){\makebox(0,0)[b]{\smash{\SetFigFont{11}{13.2}{\rmdefault}{\mddefault}{\updefault}M}}}
\put(3601,-2086){\makebox(0,0)[b]{\smash{\SetFigFont{11}{13.2}{\rmdefault}{\mddefault}{\updefault}$c_{n+1}$}}}
\put(7351,-2086){\makebox(0,0)[b]{\smash{\SetFigFont{11}{13.2}{\rmdefault}{\mddefault}{\updefault}$c_n$}}}
\put(6901,-2236){\makebox(0,0)[b]{\smash{\SetFigFont{11}{13.2}{\rmdefault}{\mddefault}{\updefault}$c_{n+1}$}}}
\end{picture}

%% file: spirals.pstex_t
\begin{picture}(0,0)%
\epsfig{file=spirals.pstex}%
\end{picture}%
\setlength{\unitlength}{3552sp}%
\begingroup\makeatletter\ifx\SetFigFont\undefined%
\gdef\SetFigFont#1#2#3#4#5{%
  \reset@font\fontsize{#1}{#2pt}%
  \fontfamily{#3}\fontseries{#4}\fontshape{#5}%
  \selectfont}%
\fi\endgroup%
\begin{picture}(6541,2163)(1579,-4643)
\put(6035,-3132){\makebox(0,0)[lb]{\smash{\SetFigFont{12}{14.4}{\rmdefault}{\mddefault}{\updefault}$(c_{i-1}, c_i)$}}}
\put(5258,-2661){\makebox(0,0)[lb]{\smash{\SetFigFont{12}{14.4}{\rmdefault}{\mddefault}{\updefault}$(c_j, c_{j+1})$}}}
\put(2498,-3105){\makebox(0,0)[lb]{\smash{\SetFigFont{12}{14.4}{\rmdefault}{\mddefault}{\updefault}$(c_j, c_{j+1})$}}}
\put(1771,-2675){\makebox(0,0)[lb]{\smash{\SetFigFont{12}{14.4}{\rmdefault}{\mddefault}{\updefault}$(c_{i-1}, c_i)$}}}
\put(5853,-4225){\makebox(0,0)[lb]{\smash{\SetFigFont{12}{14.4}{\rmdefault}{\mddefault}{\updefault}$(c_{i-1}, c_i)$}}}
\put(2356,-4186){\makebox(0,0)[lb]{\smash{\SetFigFont{12}{14.4}{\rmdefault}{\mddefault}{\updefault}$(c_j, c_{j+1})$}}}
\end{picture}

%% file: mingling_unfoldable.pstex_t
\begin{picture}(0,0)%
\epsfig{file=mingling_unfoldable.pstex}%
\end{picture}%
\setlength{\unitlength}{3158sp}%
\begingroup\makeatletter\ifx\SetFigFont\undefined%
\gdef\SetFigFont#1#2#3#4#5{%
  \reset@font\fontsize{#1}{#2pt}%
  \fontfamily{#3}\fontseries{#4}\fontshape{#5}%
  \selectfont}%
\fi\endgroup%
\begin{picture}(5010,1185)(1264,-2011)
\put(4126,-2011){\makebox(0,0)[b]{\smash{\SetFigFont{10}{12.0}{\rmdefault}{\mddefault}{\updefault}M}}}
\put(5026,-2011){\makebox(0,0)[b]{\smash{\SetFigFont{10}{12.0}{\rmdefault}{\mddefault}{\updefault}M}}}
\put(1651,-1411){\makebox(0,0)[b]{\smash{\SetFigFont{10}{12.0}{\rmdefault}{\mddefault}{\updefault}Crimp}}}
\put(3226,-961){\makebox(0,0)[b]{\smash{\SetFigFont{10}{12.0}{\rmdefault}{\mddefault}{\updefault}M}}}
\put(3526,-961){\makebox(0,0)[b]{\smash{\SetFigFont{10}{12.0}{\rmdefault}{\mddefault}{\updefault}V}}}
\put(4126,-961){\makebox(0,0)[b]{\smash{\SetFigFont{10}{12.0}{\rmdefault}{\mddefault}{\updefault}M}}}
\put(5026,-961){\makebox(0,0)[b]{\smash{\SetFigFont{10}{12.0}{\rmdefault}{\mddefault}{\updefault}M}}}
\end{picture}

%% file: moving_stuff.pstex_t
\begin{picture}(0,0)%
\includegraphics{moving_stuff.pstex}%
\end{picture}%
\setlength{\unitlength}{3355sp}%
\begingroup\makeatletter\ifx\SetFigFont\undefined%
\gdef\SetFigFont#1#2#3#4#5{%
  \reset@font\fontsize{#1}{#2pt}%
  \fontfamily{#3}\fontseries{#4}\fontshape{#5}%
  \selectfont}%
\fi\endgroup%
\begin{picture}(9393,1899)(1192,-9373)
\put(2277,-8854){\makebox(0,0)[rb]{\smash{{\SetFigFont{10}{12.0}{\rmdefault}{\mddefault}{\updefault}{\color[rgb]{0,0,0}$c_{i{+}1}$}%
}}}}
\put(7576,-8784){\makebox(0,0)[lb]{\smash{{\SetFigFont{10}{12.0}{\rmdefault}{\mddefault}{\updefault}{\color[rgb]{0,0,0}$c_{i{+}1}$}%
}}}}
\put(9376,-8634){\makebox(0,0)[rb]{\smash{{\SetFigFont{10}{12.0}{\rmdefault}{\mddefault}{\updefault}{\color[rgb]{0,0,0}$c_i$}%
}}}}
\put(9601,-8836){\makebox(0,0)[lb]{\smash{{\SetFigFont{10}{12.0}{\rmdefault}{\mddefault}{\updefault}{\color[rgb]{0,0,0}$c_{i{+}2}$}%
}}}}
\put(7351,-8686){\makebox(0,0)[rb]{\smash{{\SetFigFont{10}{12.0}{\rmdefault}{\mddefault}{\updefault}{\color[rgb]{0,0,0}$c_{i{-}1}$}%
}}}}
\put(2251,-8011){\makebox(0,0)[rb]{\smash{{\SetFigFont{10}{12.0}{\rmdefault}{\mddefault}{\updefault}{\color[rgb]{0,0,0}$c_{i{-}1}$}%
}}}}
\put(4351,-9211){\makebox(0,0)[lb]{\smash{{\SetFigFont{10}{12.0}{\rmdefault}{\mddefault}{\updefault}{\color[rgb]{0,0,0}$c_{i{+}2}$}%
}}}}
\put(4318,-8161){\makebox(0,0)[lb]{\smash{{\SetFigFont{10}{12.0}{\rmdefault}{\mddefault}{\updefault}{\color[rgb]{0,0,0}$c_i$}%
}}}}
\end{picture}%

%% file: 2dhard.pstex_t
\begin{picture}(0,0)%
\includegraphics{2dhard.pstex}%
\end{picture}%
\setlength{\unitlength}{2565sp}%
\begingroup\makeatletter\ifx\SetFigFont\undefined%
\gdef\SetFigFont#1#2#3#4#5{%
  \reset@font\fontsize{#1}{#2pt}%
  \fontfamily{#3}\fontseries{#4}\fontshape{#5}%
  \selectfont}%
\fi\endgroup%
\begin{picture}(8488,9024)(225,-8323)
\put(4801,-3136){\makebox(0,0)[b]{\smash{{\SetFigFont{8}{9.6}{\rmdefault}{\mddefault}{\updefault}{\color[rgb]{0,0,0}$v_0$}%
}}}}
\put(2926,-3736){\makebox(0,0)[lb]{\smash{{\SetFigFont{8}{9.6}{\rmdefault}{\mddefault}{\updefault}{\color[rgb]{0,0,0}$v_{n+1}$}%
}}}}
\put(2476,-5086){\makebox(0,0)[lb]{\smash{{\SetFigFont{8}{9.6}{\rmdefault}{\mddefault}{\updefault}{\color[rgb]{0,0,0}$v_{n+2}$}%
}}}}
\put(5551,-2011){\makebox(0,0)[lb]{\smash{{\SetFigFont{8}{9.6}{\rmdefault}{\mddefault}{\updefault}{\color[rgb]{0,0,0}$L$}%
}}}}
\put(6751,-1411){\makebox(0,0)[lb]{\smash{{\SetFigFont{8}{9.6}{\rmdefault}{\mddefault}{\updefault}{\color[rgb]{0,0,0}$2L$}%
}}}}
\put(4876,-1411){\makebox(0,0)[lb]{\smash{{\SetFigFont{8}{9.6}{\rmdefault}{\mddefault}{\updefault}{\color[rgb]{0,0,0}$P_4$}%
}}}}
\put(4501,-1186){\makebox(0,0)[b]{\smash{{\SetFigFont{8}{9.6}{\rmdefault}{\mddefault}{\updefault}{\color[rgb]{0,0,0}$v_1$}%
}}}}
\put(3376,-1936){\makebox(0,0)[rb]{\smash{{\SetFigFont{8}{9.6}{\rmdefault}{\mddefault}{\updefault}{\color[rgb]{0,0,0}$v_2$}%
}}}}
\put(2926,-2686){\makebox(0,0)[rb]{\smash{{\SetFigFont{8}{9.6}{\rmdefault}{\mddefault}{\updefault}{\color[rgb]{0,0,0}$v_3$}%
}}}}
\put(2476,-3286){\makebox(0,0)[rb]{\smash{{\SetFigFont{8}{9.6}{\rmdefault}{\mddefault}{\updefault}{\color[rgb]{0,0,0}$v_n$}%
}}}}
\put(4426,-2536){\makebox(0,0)[rb]{\smash{{\SetFigFont{8}{9.6}{\rmdefault}{\mddefault}{\updefault}{\color[rgb]{0,0,0}$P_3$}%
}}}}
\put(4426,-136){\makebox(0,0)[rb]{\smash{{\SetFigFont{8}{9.6}{\rmdefault}{\mddefault}{\updefault}{\color[rgb]{0,0,0}$P_0$}%
}}}}
\put(7726,-136){\makebox(0,0)[rb]{\smash{{\SetFigFont{8}{9.6}{\rmdefault}{\mddefault}{\updefault}{\color[rgb]{0,0,0}$P_1$}%
}}}}
\put(2026,-3736){\makebox(0,0)[rb]{\smash{{\SetFigFont{8}{9.6}{\rmdefault}{\mddefault}{\updefault}{\color[rgb]{0,0,0}$P_5$}%
}}}}
\put(1576,-5086){\makebox(0,0)[rb]{\smash{{\SetFigFont{8}{9.6}{\rmdefault}{\mddefault}{\updefault}{\color[rgb]{0,0,0}$P_6$}%
}}}}
\put(7726,-2536){\makebox(0,0)[rb]{\smash{{\SetFigFont{8}{9.6}{\rmdefault}{\mddefault}{\updefault}{\color[rgb]{0,0,0}$P_2$}%
}}}}
\put(8551,179){\makebox(0,0)[b]{\smash{{\SetFigFont{8}{9.6}{\rmdefault}{\mddefault}{\updefault}{\color[rgb]{0,0,0}$\epsilon$}%
}}}}
\put(826,-1636){\makebox(0,0)[rb]{\smash{{\SetFigFont{8}{9.6}{\rmdefault}{\mddefault}{\updefault}{\color[rgb]{0,0,0}$a_1$}%
}}}}
\put(826,-2311){\makebox(0,0)[rb]{\smash{{\SetFigFont{8}{9.6}{\rmdefault}{\mddefault}{\updefault}{\color[rgb]{0,0,0}$a_2$}%
}}}}
\put(826,-2986){\makebox(0,0)[rb]{\smash{{\SetFigFont{8}{9.6}{\rmdefault}{\mddefault}{\updefault}{\color[rgb]{0,0,0}$a_3$}%
}}}}
\put(826,-3511){\makebox(0,0)[rb]{\smash{{\SetFigFont{8}{9.6}{\rmdefault}{\mddefault}{\updefault}{\color[rgb]{0,0,0}$a_n$}%
}}}}
\put(826,-6436){\makebox(0,0)[rb]{\smash{{\SetFigFont{8}{9.6}{\rmdefault}{\mddefault}{\updefault}{\color[rgb]{0,0,0}$2L$}%
}}}}
\put(826,-4411){\makebox(0,0)[rb]{\smash{{\SetFigFont{8}{9.6}{\rmdefault}{\mddefault}{\updefault}{\color[rgb]{0,0,0}$L$}%
}}}}
\put(1576,-7786){\makebox(0,0)[rb]{\smash{{\SetFigFont{8}{9.6}{\rmdefault}{\mddefault}{\updefault}{\color[rgb]{0,0,0}$P_7$}%
}}}}
\put(6151,-8011){\makebox(0,0)[b]{\smash{{\SetFigFont{8}{9.6}{\rmdefault}{\mddefault}{\updefault}{\color[rgb]{0,0,0}$W_2$}%
}}}}
\put(3076,-8011){\makebox(0,0)[b]{\smash{{\SetFigFont{8}{9.6}{\rmdefault}{\mddefault}{\updefault}{\color[rgb]{0,0,0}$W_1$}%
}}}}
\end{picture}%

%% file: semifolded.pstex_t
\begin{picture}(0,0)%
\includegraphics{semifolded.pstex}%
\end{picture}%
\setlength{\unitlength}{2565sp}%
\begingroup\makeatletter\ifx\SetFigFont\undefined%
\gdef\SetFigFont#1#2#3#4#5{%
  \reset@font\fontsize{#1}{#2pt}%
  \fontfamily{#3}\fontseries{#4}\fontshape{#5}%
  \selectfont}%
\fi\endgroup%
\begin{picture}(8488,4524)(225,-3823)
\put(4501,-1186){\makebox(0,0)[b]{\smash{{\SetFigFont{8}{9.6}{\rmdefault}{\mddefault}{\updefault}{\color[rgb]{0,0,0}$v_1$}%
}}}}
\put(6151,-3511){\makebox(0,0)[b]{\smash{{\SetFigFont{8}{9.6}{\rmdefault}{\mddefault}{\updefault}{\color[rgb]{0,0,0}$W_2$}%
}}}}
\put(3076,-3511){\makebox(0,0)[b]{\smash{{\SetFigFont{8}{9.6}{\rmdefault}{\mddefault}{\updefault}{\color[rgb]{0,0,0}$W_1$}%
}}}}
\put(4801,-3136){\makebox(0,0)[b]{\smash{{\SetFigFont{8}{9.6}{\rmdefault}{\mddefault}{\updefault}{\color[rgb]{0,0,0}$v_0$}%
}}}}
\put(976,-661){\makebox(0,0)[lb]{\smash{{\SetFigFont{8}{9.6}{\rmdefault}{\mddefault}{\updefault}{\color[rgb]{0,0,0}$L$}%
}}}}
\put(676,-1336){\makebox(0,0)[rb]{\smash{{\SetFigFont{8}{9.6}{\rmdefault}{\mddefault}{\updefault}{\color[rgb]{0,0,0}$2L$}%
}}}}
\put(4876,-1411){\makebox(0,0)[lb]{\smash{{\SetFigFont{8}{9.6}{\rmdefault}{\mddefault}{\updefault}{\color[rgb]{0,0,0}$P_4$}%
}}}}
\put(2851,-811){\makebox(0,0)[lb]{\smash{{\SetFigFont{8}{9.6}{\rmdefault}{\mddefault}{\updefault}{\color[rgb]{0,0,0}$v_{n+1}$}%
}}}}
\put(1576,-2686){\makebox(0,0)[rb]{\smash{{\SetFigFont{8}{9.6}{\rmdefault}{\mddefault}{\updefault}{\color[rgb]{0,0,0}$P_7$}%
}}}}
\put(1576, 14){\makebox(0,0)[rb]{\smash{{\SetFigFont{8}{9.6}{\rmdefault}{\mddefault}{\updefault}{\color[rgb]{0,0,0}$P_6$}%
}}}}
\put(4426,-136){\makebox(0,0)[rb]{\smash{{\SetFigFont{8}{9.6}{\rmdefault}{\mddefault}{\updefault}{\color[rgb]{0,0,0}$P_0$}%
}}}}
\put(2326,-2011){\makebox(0,0)[b]{\smash{{\SetFigFont{8}{9.6}{\rmdefault}{\mddefault}{\updefault}{\color[rgb]{0,0,0}$P_5$}%
}}}}
\put(4426,-2536){\makebox(0,0)[rb]{\smash{{\SetFigFont{8}{9.6}{\rmdefault}{\mddefault}{\updefault}{\color[rgb]{0,0,0}$P_3$}%
}}}}
\put(5476,-2011){\makebox(0,0)[lb]{\smash{{\SetFigFont{8}{9.6}{\rmdefault}{\mddefault}{\updefault}{\color[rgb]{0,0,0}$L$}%
}}}}
\put(7126,-1336){\makebox(0,0)[rb]{\smash{{\SetFigFont{8}{9.6}{\rmdefault}{\mddefault}{\updefault}{\color[rgb]{0,0,0}$2L$}%
}}}}
\put(7726,-2536){\makebox(0,0)[rb]{\smash{{\SetFigFont{8}{9.6}{\rmdefault}{\mddefault}{\updefault}{\color[rgb]{0,0,0}$P_2$}%
}}}}
\put(7726,-136){\makebox(0,0)[rb]{\smash{{\SetFigFont{8}{9.6}{\rmdefault}{\mddefault}{\updefault}{\color[rgb]{0,0,0}$P_1$}%
}}}}
\put(8551,179){\makebox(0,0)[b]{\smash{{\SetFigFont{8}{9.6}{\rmdefault}{\mddefault}{\updefault}{\color[rgb]{0,0,0}$\epsilon$}%
}}}}
\end{picture}%

%% file: noassign.pstex_t
\begin{picture}(0,0)%
\includegraphics{noassign.pstex}%
\end{picture}%
\setlength{\unitlength}{2565sp}%
\begingroup\makeatletter\ifx\SetFigFont\undefined%
\gdef\SetFigFont#1#2#3#4#5{%
  \reset@font\fontsize{#1}{#2pt}%
  \fontfamily{#3}\fontseries{#4}\fontshape{#5}%
  \selectfont}%
\fi\endgroup%
\begin{picture}(8349,9024)(364,-8323)
\put(6151,-8011){\makebox(0,0)[b]{\smash{{\SetFigFont{8}{9.6}{\rmdefault}{\mddefault}{\updefault}{\color[rgb]{0,0,0}$W_2$}%
}}}}
\put(4501,-1711){\makebox(0,0)[rb]{\smash{{\SetFigFont{8}{9.6}{\rmdefault}{\mddefault}{\updefault}{\color[rgb]{0,0,0}$v_1$}%
}}}}
\put(3076,-8011){\makebox(0,0)[b]{\smash{{\SetFigFont{8}{9.6}{\rmdefault}{\mddefault}{\updefault}{\color[rgb]{0,0,0}$W_1$}%
}}}}
\put(2926,-3736){\makebox(0,0)[lb]{\smash{{\SetFigFont{8}{9.6}{\rmdefault}{\mddefault}{\updefault}{\color[rgb]{0,0,0}$v_{n+1}$}%
}}}}
\put(2476,-5086){\makebox(0,0)[lb]{\smash{{\SetFigFont{8}{9.6}{\rmdefault}{\mddefault}{\updefault}{\color[rgb]{0,0,0}$v_{n+2}$}%
}}}}
\put(1576,-5086){\makebox(0,0)[rb]{\smash{{\SetFigFont{8}{9.6}{\rmdefault}{\mddefault}{\updefault}{\color[rgb]{0,0,0}$P_6$}%
}}}}
\put(2026,-3736){\makebox(0,0)[rb]{\smash{{\SetFigFont{8}{9.6}{\rmdefault}{\mddefault}{\updefault}{\color[rgb]{0,0,0}$P_5$}%
}}}}
\put(8551,179){\makebox(0,0)[b]{\smash{{\SetFigFont{8}{9.6}{\rmdefault}{\mddefault}{\updefault}{\color[rgb]{0,0,0}$\epsilon$}%
}}}}
\put(826,-1636){\makebox(0,0)[rb]{\smash{{\SetFigFont{8}{9.6}{\rmdefault}{\mddefault}{\updefault}{\color[rgb]{0,0,0}$a_1$}%
}}}}
\put(826,-2311){\makebox(0,0)[rb]{\smash{{\SetFigFont{8}{9.6}{\rmdefault}{\mddefault}{\updefault}{\color[rgb]{0,0,0}$a_2$}%
}}}}
\put(826,-2986){\makebox(0,0)[rb]{\smash{{\SetFigFont{8}{9.6}{\rmdefault}{\mddefault}{\updefault}{\color[rgb]{0,0,0}$a_3$}%
}}}}
\put(826,-3511){\makebox(0,0)[rb]{\smash{{\SetFigFont{8}{9.6}{\rmdefault}{\mddefault}{\updefault}{\color[rgb]{0,0,0}$a_n$}%
}}}}
\put(826,-4411){\makebox(0,0)[rb]{\smash{{\SetFigFont{8}{9.6}{\rmdefault}{\mddefault}{\updefault}{\color[rgb]{0,0,0}$L$}%
}}}}
\put(826,-6436){\makebox(0,0)[rb]{\smash{{\SetFigFont{8}{9.6}{\rmdefault}{\mddefault}{\updefault}{\color[rgb]{0,0,0}$2L$}%
}}}}
\put(1576,-7786){\makebox(0,0)[rb]{\smash{{\SetFigFont{8}{9.6}{\rmdefault}{\mddefault}{\updefault}{\color[rgb]{0,0,0}$P_7$}%
}}}}
\put(2476,-3286){\makebox(0,0)[rb]{\smash{{\SetFigFont{8}{9.6}{\rmdefault}{\mddefault}{\updefault}{\color[rgb]{0,0,0}$v_n$}%
}}}}
\put(2926,-2686){\makebox(0,0)[rb]{\smash{{\SetFigFont{8}{9.6}{\rmdefault}{\mddefault}{\updefault}{\color[rgb]{0,0,0}$v_3$}%
}}}}
\put(3376,-1936){\makebox(0,0)[rb]{\smash{{\SetFigFont{8}{9.6}{\rmdefault}{\mddefault}{\updefault}{\color[rgb]{0,0,0}$v_2$}%
}}}}
\put(4801,-3136){\makebox(0,0)[b]{\smash{{\SetFigFont{8}{9.6}{\rmdefault}{\mddefault}{\updefault}{\color[rgb]{0,0,0}$v_0$}%
}}}}
\put(4426,-2686){\makebox(0,0)[rb]{\smash{{\SetFigFont{8}{9.6}{\rmdefault}{\mddefault}{\updefault}{\color[rgb]{0,0,0}$P_3$}%
}}}}
\put(4876,-1336){\makebox(0,0)[lb]{\smash{{\SetFigFont{8}{9.6}{\rmdefault}{\mddefault}{\updefault}{\color[rgb]{0,0,0}$P_4$}%
}}}}
\put(5476,-2011){\makebox(0,0)[lb]{\smash{{\SetFigFont{8}{9.6}{\rmdefault}{\mddefault}{\updefault}{\color[rgb]{0,0,0}$L$}%
}}}}
\put(7126,-1336){\makebox(0,0)[rb]{\smash{{\SetFigFont{8}{9.6}{\rmdefault}{\mddefault}{\updefault}{\color[rgb]{0,0,0}$2L$}%
}}}}
\put(7726,-136){\makebox(0,0)[rb]{\smash{{\SetFigFont{8}{9.6}{\rmdefault}{\mddefault}{\updefault}{\color[rgb]{0,0,0}$P_1$}%
}}}}
\put(7726,-2536){\makebox(0,0)[rb]{\smash{{\SetFigFont{8}{9.6}{\rmdefault}{\mddefault}{\updefault}{\color[rgb]{0,0,0}$P_2$}%
}}}}
\put(4426, 14){\makebox(0,0)[rb]{\smash{{\SetFigFont{8}{9.6}{\rmdefault}{\mddefault}{\updefault}{\color[rgb]{0,0,0}$P_0$}%
}}}}
\end{picture}%

%% file: zoom.pstex_t
\begin{picture}(0,0)%
\epsfig{file=zoom.pstex}%
\end{picture}%
\setlength{\unitlength}{2960sp}%
\begingroup\makeatletter\ifx\SetFigFont\undefined%
\gdef\SetFigFont#1#2#3#4#5{%
  \reset@font\fontsize{#1}{#2pt}%
  \fontfamily{#3}\fontseries{#4}\fontshape{#5}%
  \selectfont}%
\fi\endgroup%
\begin{picture}(5803,3742)(289,-3181)
\put(4280,-1231){\makebox(0,0)[lb]{\smash{\SetFigFont{9}{10.8}{\rmdefault}{\mddefault}{\updefault}$v_1$}}}
\put(1576,-3136){\makebox(0,0)[lb]{\smash{\SetFigFont{9}{10.8}{\rmdefault}{\mddefault}{\updefault}$v_0$}}}
\put(526,-1186){\makebox(0,0)[lb]{\smash{\SetFigFont{9}{10.8}{\rmdefault}{\mddefault}{\updefault}$v_1$}}}
\put(5180,-3181){\makebox(0,0)[lb]{\smash{\SetFigFont{9}{10.8}{\rmdefault}{\mddefault}{\updefault}$v_0$}}}
\put(5026,-965){\makebox(0,0)[lb]{\smash{\SetFigFont{9}{10.8}{\rmdefault}{\mddefault}{\updefault}$c_0$}}}
\put(5483,-2120){\makebox(0,0)[lb]{\smash{\SetFigFont{9}{10.8}{\rmdefault}{\mddefault}{\updefault}$c_1$}}}
\put(2926,-1730){\makebox(0,0)[lb]{\smash{\SetFigFont{9}{10.8}{\rmdefault}{\mddefault}{\updefault}$c_2$}}}
\end{picture}

%% file: crimp_crossing.pstex_t
\begin{picture}(0,0)%
\includegraphics{crimp_crossing.pstex}%
\end{picture}%
\setlength{\unitlength}{1579sp}%
\begingroup\makeatletter\ifx\SetFigFont\undefined%
\gdef\SetFigFont#1#2#3#4#5{%
  \reset@font\fontsize{#1}{#2pt}%
  \fontfamily{#3}\fontseries{#4}\fontshape{#5}%
  \selectfont}%
\fi\endgroup%
\begin{picture}(18535,3721)(204,-3160)
\put(17101,-3061){\makebox(0,0)[b]{\smash{{\SetFigFont{9}{10.8}{\rmdefault}{\mddefault}{\updefault}{\color[rgb]{0,0,0}Both crimps unfolded}%
}}}}
\put(1651,-3061){\makebox(0,0)[b]{\smash{{\SetFigFont{9}{10.8}{\rmdefault}{\mddefault}{\updefault}{\color[rgb]{0,0,0}First crimp marked}%
}}}}
\put(5701,-3061){\makebox(0,0)[b]{\smash{{\SetFigFont{9}{10.8}{\rmdefault}{\mddefault}{\updefault}{\color[rgb]{0,0,0}First crimp folded}%
}}}}
\put(9451,-3061){\makebox(0,0)[b]{\smash{{\SetFigFont{9}{10.8}{\rmdefault}{\mddefault}{\updefault}{\color[rgb]{0,0,0}Second crimp marked}%
}}}}
\put(13201,-3061){\makebox(0,0)[b]{\smash{{\SetFigFont{9}{10.8}{\rmdefault}{\mddefault}{\updefault}{\color[rgb]{0,0,0}Second crimp folded}%
}}}}
\end{picture}%